\documentclass[pof,graphicx,longbibliography]{revtex4-1}
\usepackage{epsfig,psfrag}
\usepackage{amsmath}
\usepackage{amsfonts}
\usepackage{mathrsfs}
\usepackage{amssymb}
\usepackage{bm}
\usepackage{color}
\usepackage{natbib}
\usepackage{mdwlist}
\usepackage{xr}

\draft 

\begin{document}

\newcommand\beq{\begin{equation}}
\newcommand\eeq{\end{equation}}
\newcommand\beqa{\begin{eqnarray}}
\newcommand\eeqa{\end{eqnarray}}
\newcommand{\Sy}{{\cal S}}
\newcommand{\U}{{\cal U}}
\newcommand{\K}{{\cal K}}
\newcommand{\T}{\mathsfi {T}} 
\newcommand{\bx}{\mathbf{x}}
\newcommand{\by}{\mathbf{y}}
\newcommand{\hx}{\mathbf{ \hat{x}}}
\newcommand{\hy}{\mathbf{ \hat{y}}}
\newcommand{\hz}{\mathbf{ \hat{z}}}


\def\half{\frac{1}{2}}
\def\quart{\frac{1}{4}}
\def\d{{\rm d}}
\def\eps{\epsilon}

\title{Recurrent flow analysis in spatiotemporally chaotic 2-dimensional Kolmogorov flow}



\author{Dan Lucas}
\email[]{dan.lucas@ucd.ie}
\homepage[]{mathsci.ucd.ie/~dan}
\altaffiliation{\\now School of Mathematical Sciences, University College Dublin, Dublin, Ireland}
\affiliation{School of Mathematics, University of Bristol, University Walk, Bristol, UK}
\author{Rich R. Kerswell}
\email[]{r.r.kerswell@bris.ac.uk}
\homepage[]{www.maths.bris.ac.uk/~marrk}
\affiliation{School of Mathematics, University of Bristol, University Walk, Bristol, UK}

\date{\today}

\begin{abstract}

Motivated by recent success in the dynamical systems approach to
transitional flow, we study the efficiency and effectiveness of
extracting simple invariant sets (recurrent flows) directly from
chaotic/turbulent flows and the potential of these sets for providing
predictions of certain statistics of the flow.  Two-dimensional
Kolmogorov flow (the 2D Navier-Stokes equations with a sinusoidal body force) is
studied both over a square $[0,2\pi]^2$ torus and a rectangular torus
extended in the forcing direction. In the former case, an order of
magnitude more recurrent flows are found than previously [G.J. Chandler \& R.R. Kerswell, J. Fluid Mech. 722, 554 (2013)] and shown to give improved predictions for the
dissipation and energy pdfs of the chaos via periodic orbit theory.
Analysis of the recurrent flows shows that the energy is largely trapped in 
the smallest wavenumbers through a combination of the inverse cascade process 
and a feature of the advective nonlinearity in 2D.
Over the extended torus at low forcing amplitudes, some extracted
states mimick the statistics of the spatially-localised chaos present
surprisingly well recalling the findings of Kawahara \& Kida
 [G. Kawahara \& S. Kida, J. Fluid Mech. 449, 291 (2001)]  in low-Reynolds-number plane Couette flow. At higher forcing
amplitudes, however, success is limited highlighting the increased
dimensionality of the chaos and the need for larger data sets.
Algorithmic developments to improve the extraction procedure are
discussed.

\end{abstract}


\maketitle 

\section{Introduction}

Recent years have seen an increasing trend towards a dynamical systems
approach to the study of at least chaotic if not turbulent
flows. While this approach has already seen considerable success in
the study of transition to turbulence
\cite{Kerswell:2005ir,Eckhardt:2007ka,Eckhardt:2008jv,
  Kawahara:2012iu}, it is very much still in its infancy when applied
to fully turbulent flows
\citep{Kawahara:2012iu,Cvitanovic13,Chandler:2013}.  Here, the
basis of the approach is the idea of a turbulent flow being
represented by a point in phase space tracing out a trajectory in
time which fleetingly but repeatedly visits the neighbourhoods of
simple invariant sets (exact solutions of the governing questions)
embedded in the phase space \citep{Hopf48}. During such a visit
the flow trajectory is attracted along the (typically very high
dimensional) stable manifold of the simple invariant set before being
expelled along its (typically much lower-dimensional) unstable
manifold, the flow transiently takes on the properties of the simple
invariant set. Given enough of these invariant sets, the hope is then
that an appropriately weighted sum of all their properties could be
used to predict those of the turbulent flow. This approach has proved
fruitful for very low-dimensional hyperbolic systems
\citep{Artuso90a,Artuso90b,Cvitanovic92,Cvitanovic13} where periodic
orbit theory (POT) is used to weight the various invariant set
contributions. However, the application to very high dimensional
turbulent flows is less clear and hugely challenging (see
\citep{Chandler:2013} for a discussion).

The situation is not without hope, though, as the breakthrough
computation in 2001 of Kawahara and Kida \citep{Kawahara:2001ft}
indicated. These authors managed to isolate a single periodic orbit
embedded in the turbulent attractor in a 15,422 degree-of-freedom
simulation of small-box plane Couette flow at low Reynolds number. The
velocity statistics of this one orbit (period $\approx$ 6 eddy turnover times) were found to be very
similar to those of the turbulent state itself (see their figure
3). Encouraged by this success, subsequent work has concentrated on
trying to extract further simple invariant sets from direct numerical
simulations - a process hereafter referred to as {\it recurrent flow
  analysis} - to improve this correspondence
\citep{vanVeen:2006fm,Viswanath:2007wc,
  Viswanath:2009vu,Cvitanovic2010,Kreilos:2012bd,Willis2013,Chandler:2013}.
However progress has been slow due the many challenges
surrounding this approach starting with how best to identify simple
invariant sets - hereafter {\it recurrent flows} - from turbulent
simulations (and subsequently converge them to machine accuracy),
followed by how best to weight the contribution of each recurrent flow
in a prediction, to finally deciding how many such flows are needed to
achieve a required prediction accuracy. So far, the furthest
this approach has been pushed for the Navier-Stokes equations is in
2-dimensional Kolmogorov flow \citep{Chandler:2013} where 50 recurrent
flows were found at a relatively low level of forcing and used to {\em
  post\,}dict an array of key statistics of the (weakly) turbulent
flow. Various weighting strategies were tested alongside periodic
orbit theory with the conclusion that not enough recurrent flows had
been extracted to see the theory outperform even a simple-minded
`democratic' approach of equal weighting. Furthermore, attempts to
repeat the procedure at a higher level of forcing failed to produce
enough recurrent flows to even attempt a prediction.  A number of
reasons for this failure were discussed, but undoubtedly the most likely
reason  was that the extent
of turbulent flow data used to identify recurrent flows needed to be
much larger. One objective here is to revisit this calculation by
generating and then processing such an extended data set.

The second objective is to make a first attempt to study an extended
system where spatiotemporal behaviour is possible. Here, a further,
more fundamental challenge emerges: how far can one push this approach
when the apparent correlation length of the flow is smaller than the
domain size? Formally, nothing has changed in the sense that global
recurrent flows still exist to the full dynamical system produced by
the Navier-Stokes equations applied over the whole domain. Informally,
however, the fact that the correlation length is smaller than the
domain implies that the flow never fully `approaches' such `global'
states in phase space. Practically, this would seem to make the
identification of near-recurrences (from the turbulent flow data) good
enough to converge to exact recurrent flows considerably more
difficult. A possible way to circumvent this could be to focus on
sub-domains where coherence is identified to isolate potentially localised
recurrent flow states. However this raises fresh issues, for example what
boundary conditions to impose on the sub-domain. To start exploring
these important issues, we treat - i.e. apply {\em recurrent flow
  analysis} to - the spatiotemporal flows recently discovered in
\citep{Lucas:2014} by extending the domain of 2-dimensional Kolmogorov
flow in the forcing direction.  Here, a series of states exhibiting
spatially-localised chaos were found which present an ideal
opportunity to broach these issues. Finally, since the method to carry
out this recurrent flow analysis is still relatively new and
unsophisticated, we also take the opportunity to discuss issues
surrounding the technique and developments undertaken to improve
our procedures.

The plan of the paper is as follows. The formulation of 2D Kolmogorov flow studied
here is described in section II and exactly follows the set up described in
\citep{Chandler:2013,Lucas:2014}. The results are presented in two
sections: section III describes the square torus and $Re=60$
calculations and section IV details the computations carried out on
the various chaotic states present in the extended domain case. A
final section V discusses all the results and indicates future
directions.


\section{Formulation}

Kolmogorov flow is the name given to body-forced incompressible
viscous flow over a doubly periodic domain where the forcing is
steady and monochromatic \citep{Arnold60} so that the
governing equations are
\begin{align}
\frac{\partial \bm u^*}{\partial t^*} + \bm u^*\cdot\nabla^*\bm u^*
 +\frac{1}{\rho}\nabla^*p^* &= \nu \Delta^* \bm u^* + \chi\sin(2\pi n y^*/L_y)\bm \hat{\bx}, \\ 
\nabla^*\cdot \bm u^* &=0
\end{align}
where $\bm u = u \hat{\bx}+v \hat{\by}=(u,v)$ is the two-dimensional
velocity field, $n$ is the forcing wavenumber, $\chi$ the forcing
amplitude, $\nu$ kinematic viscosity, $p$ pressure and $\rho$ is the
density of the fluid defined over the doubly periodic domain $(x,y)
\in [0,L_x]\times[0,L_y]$. The system is naturally
non-dimensionalised with lengthscale $L_y/2\pi$ and timescale
$\sqrt{L_y/2\pi\chi}$ to give
\begin{align}
\frac{\partial \bm u}{\partial t} + \bm u\cdot\nabla\bm u +\nabla p 
&= \frac{1}{Re} \Delta \bm u + \sin n y \,\bm \hat{x}\label{NSu},\\ 
\nabla\cdot \bm u &=0
\end{align}
where we define the Reynolds number
\begin{equation}
Re := \frac{\sqrt{\chi}}{\nu}\left(\frac{L_y}{2\pi}\right)^{3/2}
\end{equation}
and take $n=4$ throughout as in \citep{Chandler:2013,Lucas:2014}.  
The equations are solved over
the torus $[0,2\pi/\alpha]\times[0,2\pi]$ where $\alpha = L_y/L_x$
defines the aspect ratio of the domain with $\alpha=1$ in
\S \ref{sect:60} and $\alpha=\quart$ in \S \ref{rectangle}.
For computational efficiency
and accuracy (\ref{NSu}) is formulated so that vorticity $\omega :=
\hz \cdot \nabla \times \bm u$ is the prognostic variable and the vorticity
equation
\begin{align}\label{NS}
\frac{\partial \omega}{\partial t} + \bm u\cdot\nabla \omega &= 
\frac{1}{Re} \Delta \omega -n \cos n y
\end{align}
is solved numerically using the GPU timestepping code presented in
\citep{Lucas:2014}. Vorticity is discretised via a Fourier-Fourier spectral expansion
with resolution $N_x\times N_y$ and dealiased by the two-thirds rule;
\begin{align}
\omega(x,y,t) = \sum^{N_x/3-1}_{k_x=0} \sum^{N_y/3-1}_{k_y=-N_y/3} 
\Omega_{k_xk_y}(t)\mathrm{e}^{\mathrm{i}\left(\alpha k_x x+k_y y\right)}.
\end{align}
A Crank-Nicolson timestepping scheme is used for the viscous terms and
Heun's method for the nonlinear and forcing terms. Typical numerical
resolutions used were 128 Fourier modes per $2\pi$ so
$(N_x,N_y)=(128,128)$ for $\alpha=1$ and $(N_x,N_y)=(512,128) $ for
$\alpha=1/4$.  Typical time steps were $dt=0.05$ at $Re=20$ and
$0.001$ at $Re=120$ with $2\times 10^6$ time steps ($dt=0.05$,
$T=10^5$) of the $512 \times 128$ grid for $\alpha=1/4$ and $Re=20$
taking 63 minutes on a 512-core NVIDIA Tesla M2090 GPU. The system has
the following symmetries:
\begin{align}
\mathcal{S}:[u,v](x,y) \rightarrow &[-u,v]\left(-x,y+\frac{\pi}{n}\right), \label{S} \\
\mathcal{R}:[u,v](x,y) \rightarrow &[-u,-v]\left(-x,-y\right), \label{R} \\
\mathcal{T}_l:[u,v](x,y) \rightarrow &[-u,v]\left(x+l,y\right), \notag  \qquad \textrm{for } 0\leq l \leq \frac{2\pi}{\alpha} \label{Tl}
\end{align}
where $\mathcal{S}$ represents the discrete shift-and-reflect symmetry
relative to the forcing, $\mathcal{R}$ rotation through $\pi$ and
$\mathcal{T}_l$ is the group of continuous translations in $x$. In
order to discuss various features of the flows considered we define
the total energy, dissipation and energy
input:
\begin{align}
E(t) &:= \half \langle \bm u^2 \rangle_A, \qquad
D(t) := \frac{1}{Re} \langle|\nabla \bm u  |^2\rangle_A, \qquad
I(t) :=  \langle u \sin n y \rangle_A
\end{align}
as diagnostic quantities
where the area average is defined as 
\[ \langle\quad\rangle_A := \frac{\alpha}{4\pi^2} \int_0^{2\pi}\int_0^{2\pi/\alpha} \d x \d y. \]
Mean flow and rms (root mean square) fluctuation velocities are defined by
\begin{align}
U(y):= \langle \bm u\cdot \hat{\bm x} \rangle_{t,x}, \qquad 
\hat{u}^2_{rms}(t) := \langle (u-U)^2 \rangle_A, \qquad
\hat{v}^2_{rms}(t) := \langle v^2 \rangle_A, \qquad
\end{align}
(initial conditions are set so the mean of $v$ is zero and
therefore will be zero for all time). The variation of the rms quantities
only with $y$ is brought out by the quantities
\begin{align}
{u}^2_{rms}(y) := \langle (u-U)^2 \rangle_{t,x}, \qquad
{v}^2_{rms}(y) := \langle v^2 \rangle_{t,x}, \qquad
\end{align}
where
\[ \langle\quad\rangle_t := \lim_{T\rightarrow \infty} \frac{1}{T} \int_0^{T} \d t, \qquad  \langle\quad\rangle_x := \frac{\alpha}{2\pi} \int_0^{2\pi/\alpha} \d x. \]
The base state, or laminar profile, is given by
\begin{equation}
\bm u_{lam} := \frac{Re}{n^2}\sin n y \, \bm{\hat x}, \qquad 
\omega_{lam} := -\frac{Re}{n} \cos n y 
\end{equation}
with its energy 
$E_{lam} := Re^2/4n^2$
and dissipation, $D_{lam} := Re/2n^2$ used for normalisation purposes.

The purpose of this work is to discuss the extraction of recurrent
flows from simulation data of a chaotic/turbulent 2D Kolmogorov
flow. To do this we follow the methodology described in section 3.2 of \citep{Chandler:2013}. 
Briefly, `near recurrences' are defined as episodes in the DNS where
\beq
{\mathcal T}_s{\mathcal S}^{2m} \omega(x,y,t+T):=
\omega(x+s,y+\half \pi m,t+T)=\omega(x,y,t)
\label{recurrence}
\eeq
`approximately' holds for some choice of the continuous shift $0
\leq s < 2 \pi$, the discrete shift $m \in \{0,1,2,\dots,n-1\}$ and
$T>0$ over $0 \leq x,y < 2\pi$.  This expression reflects the presence
of the system symmetries $\mathcal{T}_l$ and $\mathcal{S}^2$ but not
  $\mathcal{R}$ and $\mathcal{S}$ which was a choice made in
  \cite{Chandler:2013} for reasons of expediency: recurrent flows of
  period $T$ with these suppressed symmetries are captured by (\ref{recurrence})
  as recurrent flows with period $2T$.  In (\ref{recurrence}), periodic orbits
  correspond to $s=m=0$ and some period $T>0$, travelling waves (TWs)
  to $m=0$ and $s=cT$ with $T>0$ free where $c$ is the phase speed,
  equilibria have $s=m=0$ and $T$ free and relative periodic orbits
  have one or both of $s$ and $m$ not equal to zero with period
  $T>0$. Once a near-recurrent event has been found, a
  Newton-GMRES-hookstep algorithm was employed to try to converge an
  exactly recurrent flow (at least to double precision accuracy).  The
  algorithm used is as described in \citep{Chandler:2013} albeit with
  the underlying time-stepping part of the process carried out on a
  GPU card to make the procedure faster.

%
%
%
%
\begin{figure}
\begin{center}
\includegraphics[width=0.8\textwidth]{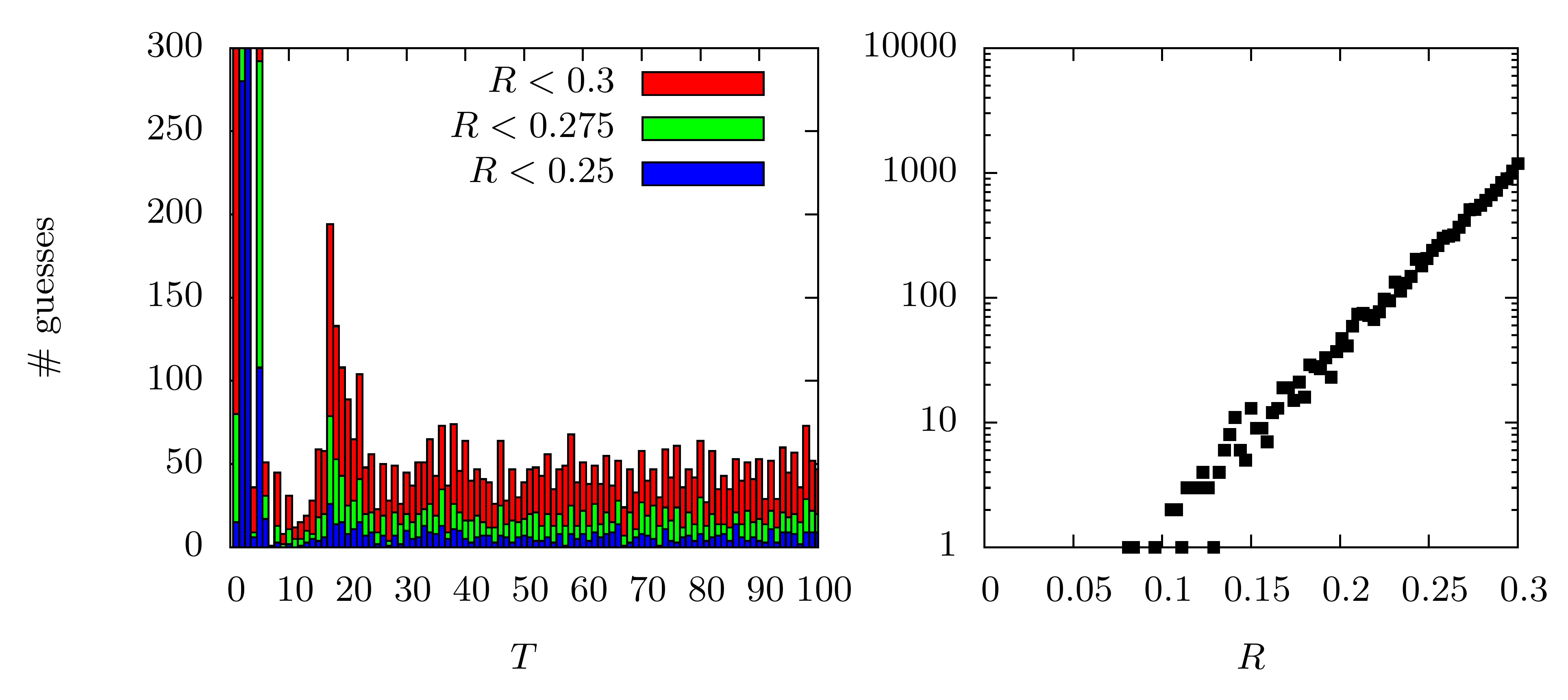}
\caption{\label{fig:guess}Distribution of recurrence guesses from the
  $Re=60$, $\alpha=1$ DNS calculation with $T=5\times 10^6$ and
  $R_{thres}=0.3$. Left plot shows a histogram of total recurrences
  determined in $\Delta T=1.0$ windows. Right plot shows the
  frequency distribution of residuals for the guesses with $R_{thres} =0.3$ on
  log scale.}
\end{center}
\end{figure}
%
%
\begin{figure}                                                             
\begin{center}\setlength{\unitlength}{1cm}              
\includegraphics[width=0.8\textwidth]{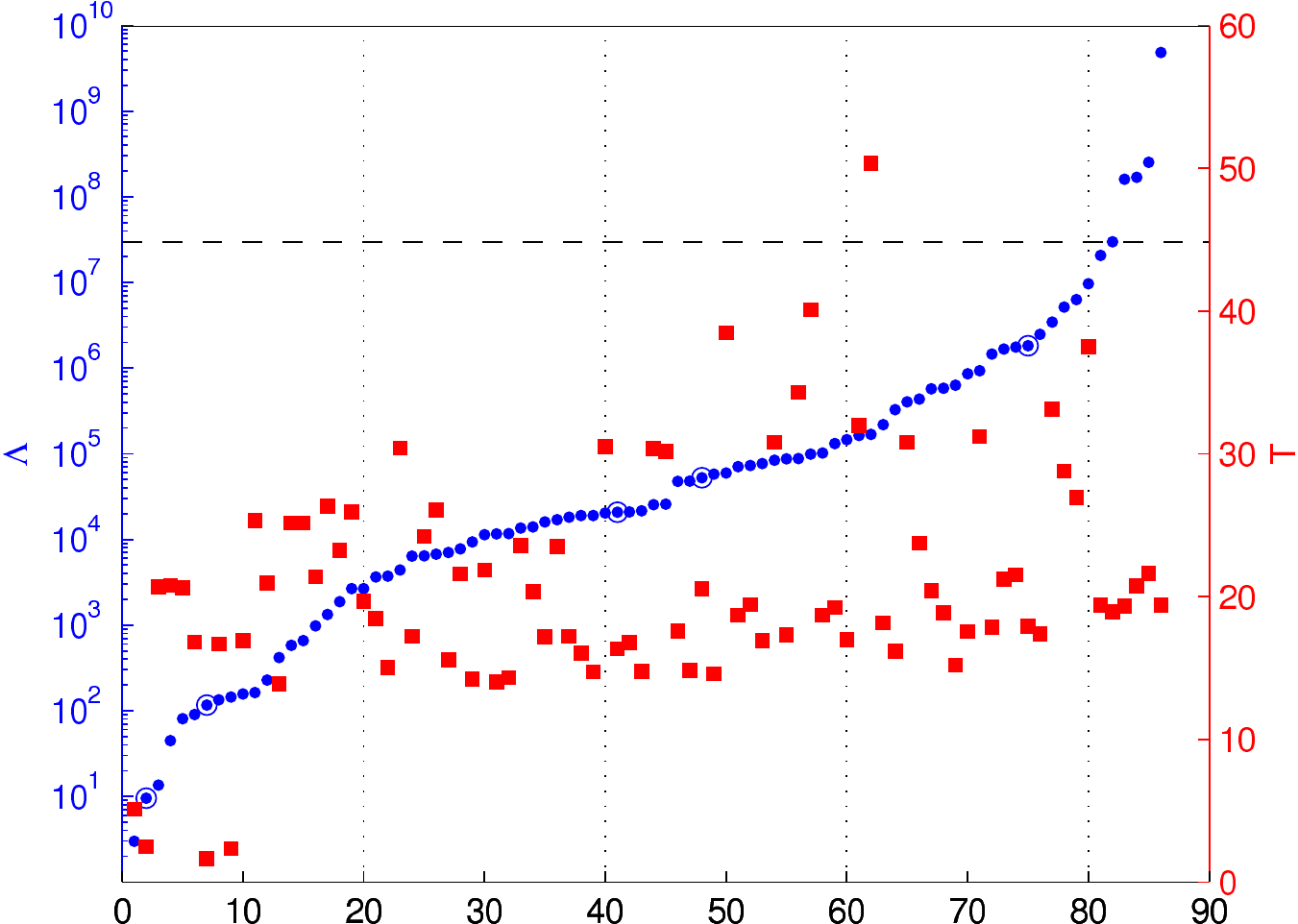}
 \end{center}                                                
\caption{The stabilities $\Lambda_i$ (see the definition in (\ref{stabcoeff})) 
of the 86 periodic and relatively
  periodic recurrent flows at $Re=60$ (blue filled circles with the original 5
  found in \cite{Chandler:2013} having an extra circle).  The black dashed line
  indicates $\Lambda_{max}(82)$, the fifth largest value, used for the
  stability cut-off.  The periods $T$ of the recurrent flows (red
  filled squares) are also shown to indicate that there is a weak
  positive correlation with stability.}
\label{UPO_60}                                                        
\end{figure}      

\section{Results: Square torus ($\alpha=1$) at $Re=60$} \label{sect:60}

In this first section of results, we revisit the square torus chaos
that was the focus of \citep{Chandler:2013}. There it was found that
at higher Reynolds numbers ($Re>40$) longer integrations were
necessary to assemble a large enough set of recurrent flows to enable
a meaningful analysis using POT.  Focussing on $Re=60$, an extended
direct numerical simulation (DNS) was conducted to $T=5\times 10^6$
which is 50 times longer than each of the 3 runs analysed in
\citep{Chandler:2013}. Near-recurrences of the flow were recorded during
the simulation to be used as initial guesses in a
Newton-GMRES-hookstep rooting-finding algorithm afterwards.

\subsection{Recurrence extraction and convergence} \label{extraction}

To expedite the extraction of guesses for recurrent flows, the
recurrence criterion of \citep{Chandler:2013} was modified
slightly. Instead of defining a close recurrence based on the
normalised difference of the full state vector, we included only the 8
largest modes in either direction;
\begin{equation}
R(t,T) :=\min_{0\leq s \leq 2\pi/\alpha}\min_{m\in 0,1,2..n-1} 
\frac{\sum\limits_{k_x=0}^{8}\sum\limits_{k_y=-8}^{8}\left | 
\Omega_{k_xk_y}(t)\mathrm{e}^{i(k_x\alpha s - 2k_ym\pi/n)}-
\Omega_{k_xk_y}(t-T) \right |^2}{\sum\limits_{k_x=0}^{8}\sum\limits_{k_y=-8}^{8}\left | 
\Omega_{k_xk_y}(t)\right |^2} < R_{thres} \label{recur}
\end{equation}
In comparisons with the `full' recurrence check ($N_x=N_y=128$ so
$\max{k_x}=\max{k_y}=42$), this reduced criterion was found to
  capture almost every near-recurrence but crucially gave a huge
  computational saving: there was a $~14$ times speed up compared to the
  full state vector criterion (see Table \ref{gss_table} in Appendix A
  and the discussion there for other recurrence criteria).  

Setting the residual threshold $R_{thres}=0.3$, the DNS calculation
yielded 11,934 guesses for when the flow appeared to repeat; figure
\ref{fig:guess} shows the distribution of these guesses over period,
$T$, and residual, $R$. The exponential increase in the number of
guesses with increasing $R$ is probably a cumulative effect since
recurrences discovered with a small residual are also likely to be
visited `less closely', i.e. with a larger residual, at about the same
time.  The variation with period $T$ is marked by a large skew toward
low periods, i.e. $T<5.0$, which reflect close visits to unstable
steady and travelling wave states. For larger periods
($T>20$), there is little variation in the numbers of guesses found. 
This is somewhat counterintuitive since it should become less probable
that the trajectory shadows a recurrent flow the longer its period
if the leading Lyapunov exponent is largely period-independent.

%
%
\begin{figure}
\begin{center}
\includegraphics[width=0.8\textwidth]{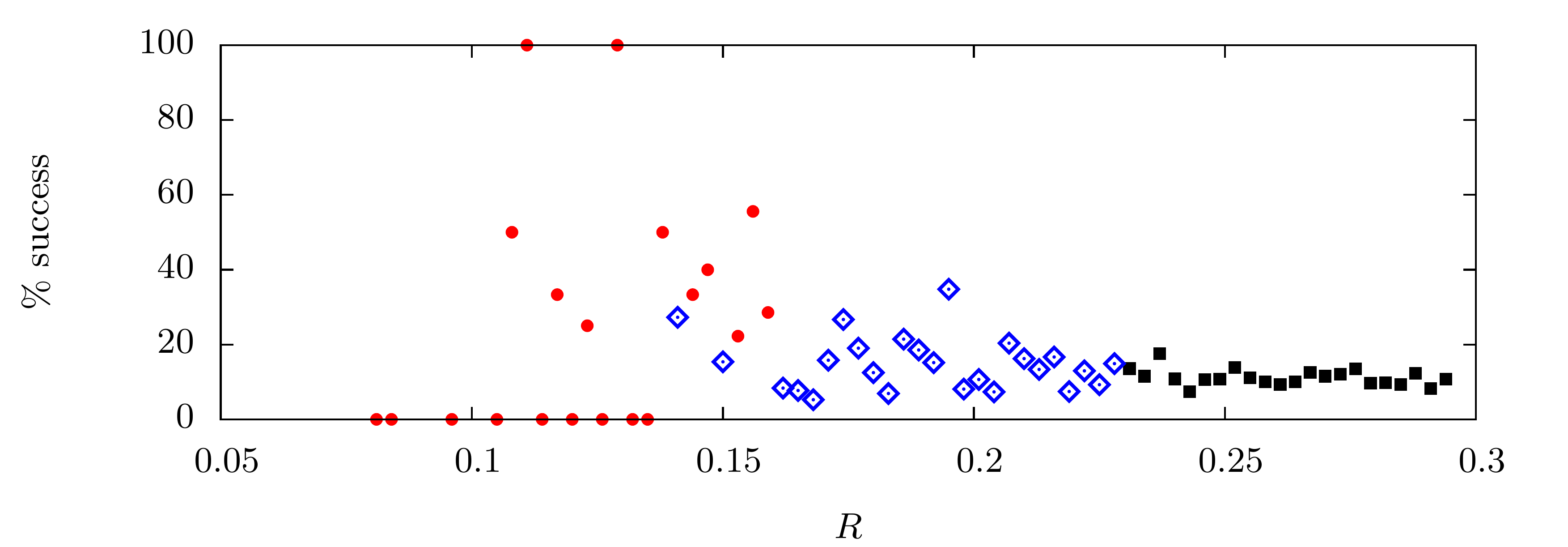}
\caption{\label{R_succ}Figure showing the convergence success rate for
  the guesses at $Re=60$ with a given starting residual (see figure
  \ref{fig:guess}). Red circles indicate where the total guesses
  are less than 10, blue diamonds have total guesses less than 100
  (and more than 10) and black squares have over 100 guesses. Note
  increased variability at small $R$ where the number of guesses is
  small and the trend toward a fixed success rate when the sample size
  increases with larger $R$.}
\end{center}
\end{figure}

From these guesses we converged 81 unique recurrent flows: see Table
\ref{UPOs} in appendix B and figure \ref{UPO_60}. At low periods, there was a large
repetition of the converged solutions already found in
\cite{Chandler:2013} ($E1, T1, T3, T4, R7, R8$ and symmetry group
permutations thereof) and so only about a third of the guesses with
$T<5.0$ were processed. Likewise for $T>60.0$, no recurrences with
$R<0.25$ converged so we focussed computational resources on periods
$0.5 < T <60.0$ and skipped recurrences with $R>0.25$ (this filtering
still meant attempts were made to converge 5,120 of the total 11,934
guesses). Figure \ref{R_succ} shows the success rate of converging
near-recurrence guesses into exactly recurrent flows as a function of
the starting residual.  As is expected, there is a negative
correlation but it is weak: for example the chances of converging a
guess settle to about $10\%$ for $R \gtrsim 0.2$.

%
%

\subsection{Recurrent processes}


Previous work \citep{Hamilton:1995,Waleffe:1997ia} in wall-bounded shear flows has
highlighted how the physical processes which combine to sustain recurrent flows (exact
coherent structures) can be exactly those which underpin the turbulent
state. Here we examine the recurrent flows found to gain some
understanding of turbulent 2D Kolmogorov flow. While the latter never
precisely repeats, periodic orbits embedded within the turbulence
represent an exactly closed cycle of dynamical processes which can be
scrutinised in great detail.

\begin{figure}
\begin{center}
\includegraphics[width=0.8\textwidth]{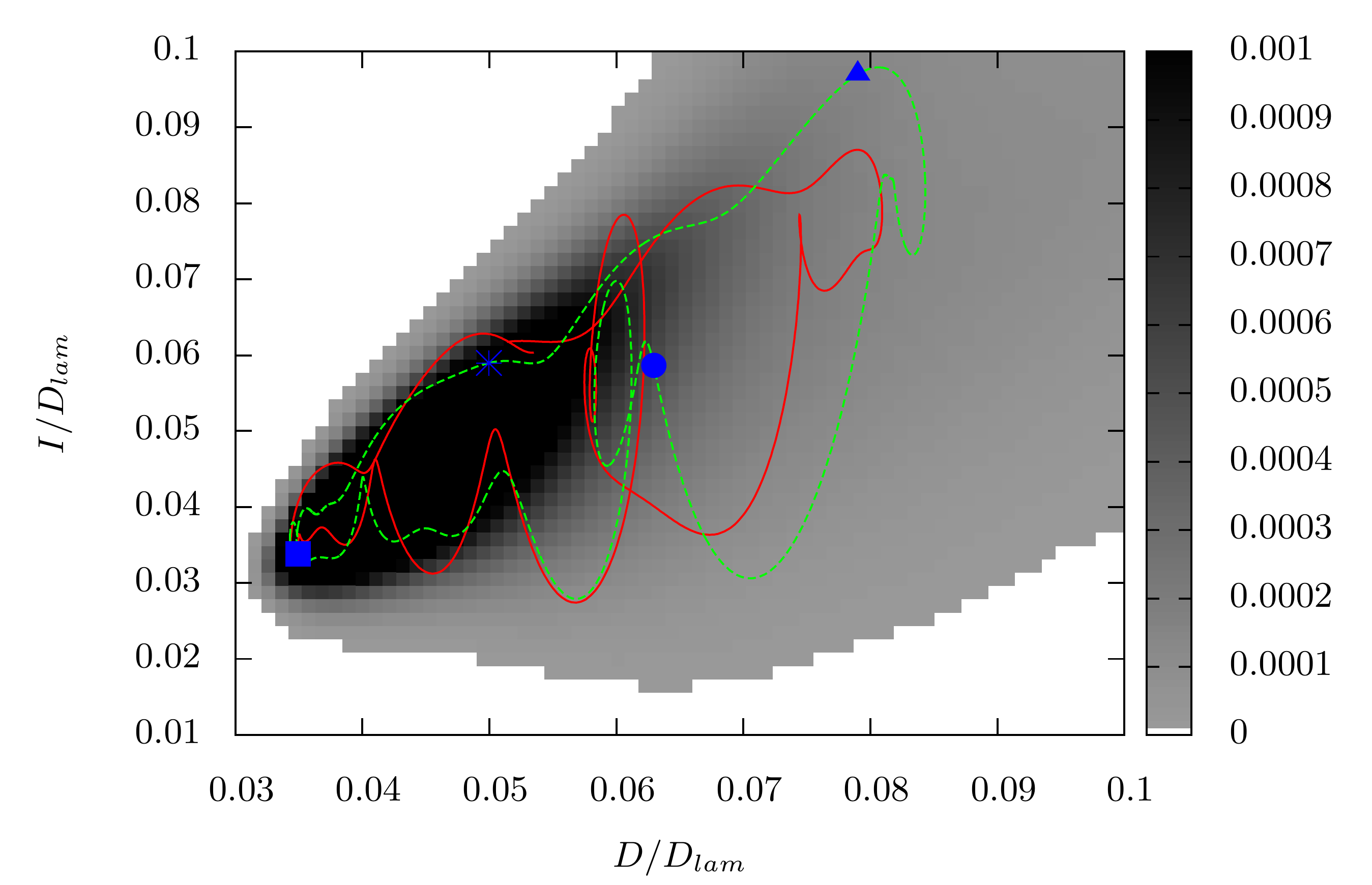}
\caption{\label{fig:DI_37} Projection of the flow onto the plane
  $I/D_{lam}$-vs-$D/D_{lam}$, greyscale showing the p.d.f. of the
  turbulent attractor, green shows UPO 37 ( $T=19.334$ and $s=0.375$)
  and red the nearby segment of DNS which constituted the initial
  guess which converged to UP0 37. Symbols correspond to the frames in
  figure \ref{UPO37_vort}, $t=0$ star, $t=3.99$ triangle, $t=7.98$
  circle and $t=15.01$ square.}
\end{center}
\end{figure}

\begin{figure}
\begin{center}
\includegraphics[width=0.8\textwidth]{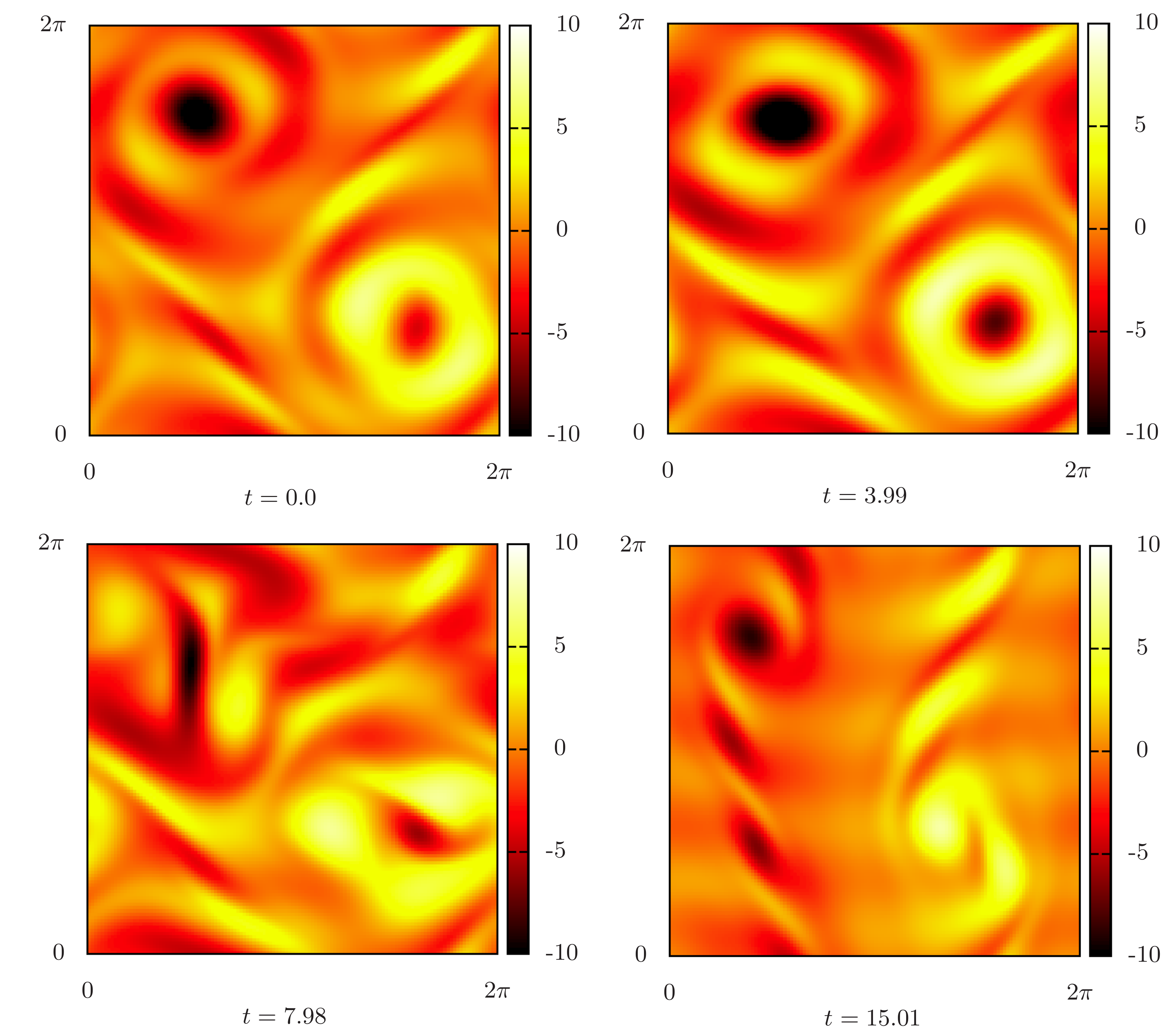}
\caption{\label{UPO37_vort}Vorticity fields for UPO 37 ($T=19.334$ and
  $s=0.375$): the shading is dark/red ($\omega<0$) through to white/light ($\omega>0$). 
	From top left to bottom right: $t=0.0$ shows the
  emergence of large scale vortices, $t=3.99$ large amplitude domain
  scale vortices (note the positive vortex is hollow due to the
  modulation of the underlying forcing), $t=7.98$ shows the generation
  of filaments and high shear associated with the destruction of the
  vortices and finally $t=15.01$ the low amplitude state signifying
  the beginning of the vortex growth phase (note this snapshot is
  indicative of the secondary kink states discovered in
  \citep{Lucas:2014}).}
\end{center}
\end{figure}

Periodic orbits with the largest range of dissipation provide the best
sampling of the chaotic attractor and the relative periodic orbit with
$T=19.33$ (UPO 37 in Table \ref{UPOs}) is a typical example.
Figure \ref{fig:DI_37} shows the projection onto the $I$-$D$ plane of
UPO 37, the segment of DNS yielding the guess which converged to UPO
37 and the underlying p.d.f of the turbulent state. A visual
inspection of the vorticity field of this orbit (figure
\ref{UPO37_vort}) gives some suggestion of a 2D inverse cascade of
energy demonstrated by a growth of (two oppositely signed) large scale
vortices (modulated by the underlying forcing, top two frames in
figure \ref{UPO37_vort}; $t=0$ and $t=3.99$). When the vortices have
grown to some critical size, an instability is triggered brought about
by their mutual interaction ($t=7.98$ in figure \ref{UPO37_vort}),
leading to strong filamentation (a direct enstrophy cascade) and
dissipation at large wavenumbers. Thereafter the mean flow reenergises the
resultant low amplitude state ($t=15.01$ in figure \ref{UPO37_vort})
establishing the `inverse cascade process' (there is only a small gap 
between the forcing wavenumber $\bm{k}_f=(0,4)$ and the
fundamental wavenumbers of the largest scales, $\bm{k}=(1,0)$ and
$(0,1)\,$).


\begin{figure}
\begin{center}
\includegraphics[width=0.8\textwidth]{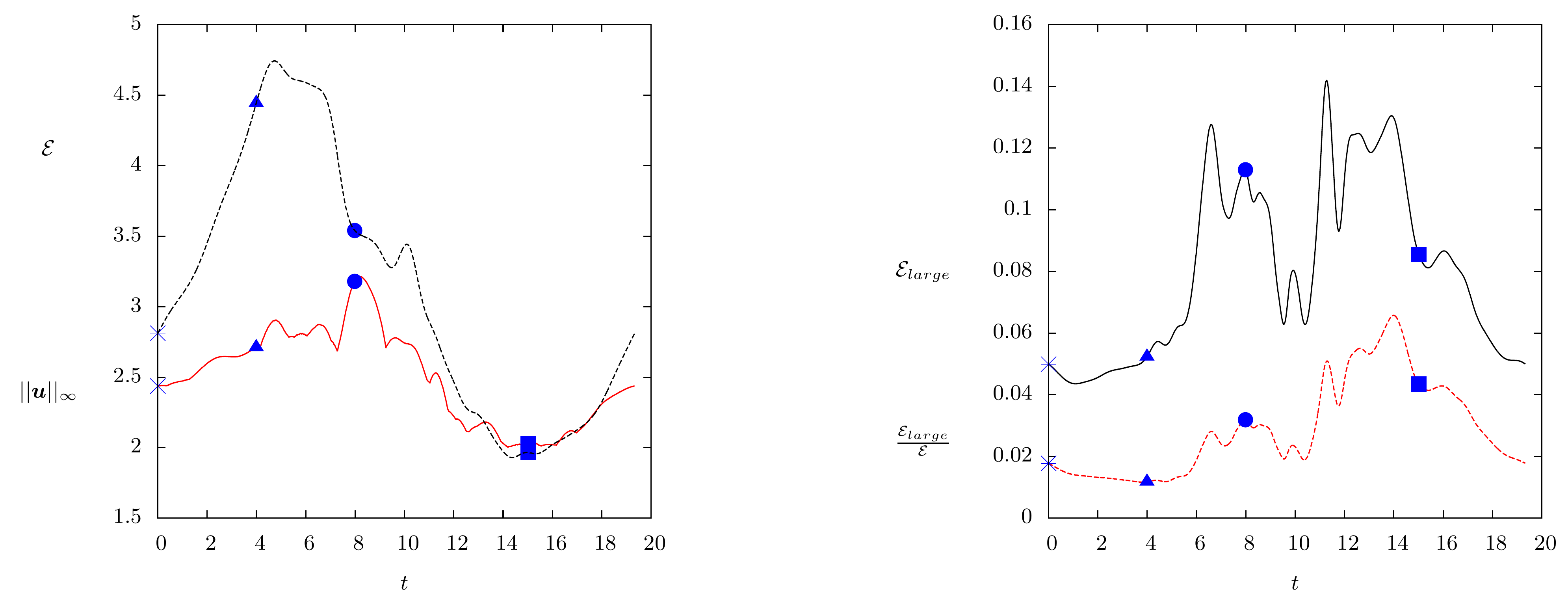}
\caption{\label{fig:UPO37_ens} Left plot shows the variation of total
  enstrophy ($\mathcal{E}$, black dashed) and maximum velocity ($||\bm
  u||_\infty$, red solid) across the period of UPO 37, right
  the proportion of enstrophy in the largest modes
  ($\mathcal{E}_{large}/\mathcal{E}$, red dashed) and the unscaled
  enstrophy in $k>32$ ($\mathcal{E}_{large}$, black solid) for the same
  orbit. One can see the build up and release of total enstrophy/flow
  amplitude as the vortices grow and then are destroyed. Similarly one
  can see some evidence of the cascade of enstrophy to small scales
  via the increase of enstrophy at high $k$. Symbols correspond to the
  frames in figure \ref{UPO37_vort}, $t=0$ star, $t=3.99$ triangle,
  $t=7.98$ circle and $t=15.01$ square.}
\end{center}
\end{figure}

To quantify this behaviour we consider a Fourier decomposition of the
flow and define the total enstrophy and enstrophy in the largest
wavenumbers as
\begin{equation}
\mathcal{E}(t) = \sum\limits_{k} |\Omega_{k_xk_y}|^2, 
\qquad \mathcal{E}_{large}(t) = \sum\limits_{k^2>32} |\Omega_{k_xk_y}|^2
\end{equation}
(note here that total enstrophy is synonymous with dissipation rate
$D(t)$).  The threshold $k^2>32$ ($=2n^2$ where $n$ is the forcing
wavenumber) was chosen as being sufficiently distant in wavenumber
space from the forcing scale to signify small scales produced by a
cascade process. Figure \ref{fig:UPO37_ens} (left) shows that the 
total enstrophy serves as a proxy for the
flow amplitude $||\bm u||_\infty$. $\mathcal{E}_{large}$
(right plot) indicates that at the large amplitude, large scale state
($t=3.99$, figure \ref{UPO37_vort}) the proportion of enstrophy in the
smallest scales has a minimum, whereas the maximum of small-scale
enstrophy immediately precedes the low amplitude (kink) state
($t=15.01$, figure \ref{UPO37_vort}) and follows the vortex breakdown
at $t=7.98$ (figure \ref{UPO37_vort}).

\begin{figure}
\begin{center}
\includegraphics[width=0.8\textwidth]{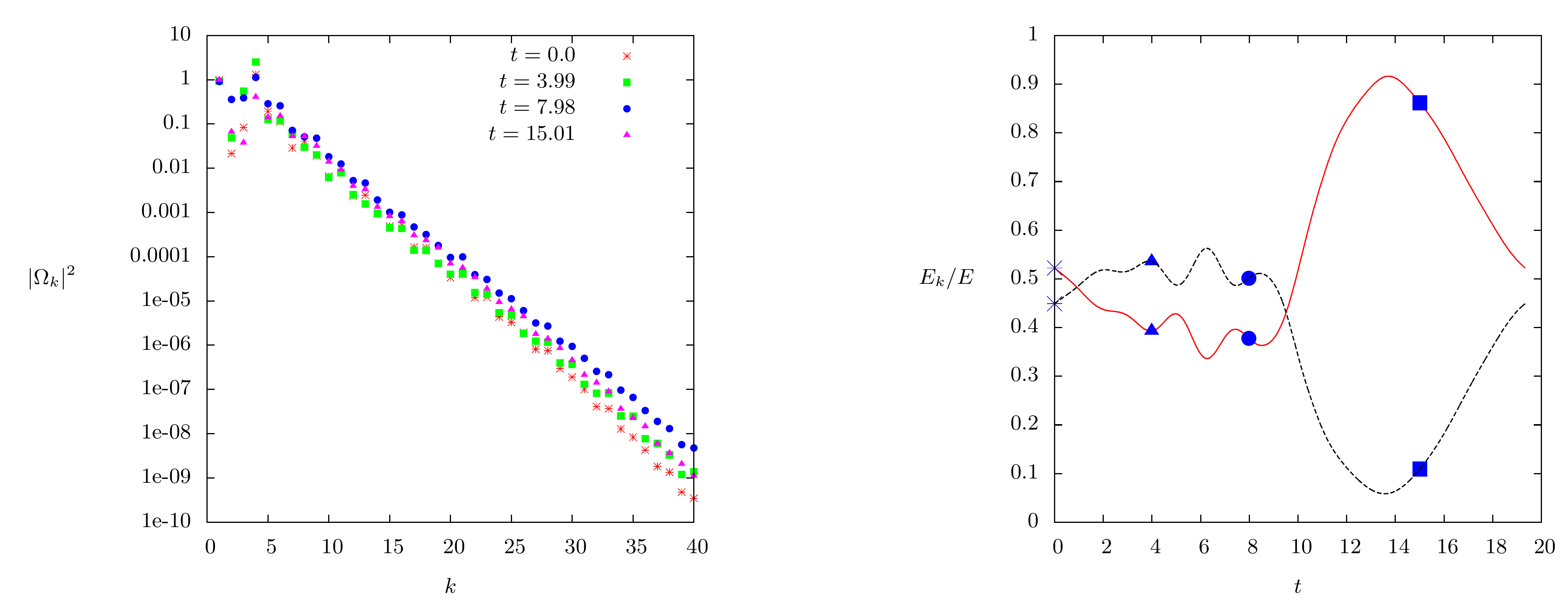}
\caption{\label{fig:UPO37_k} Figure showing left, 1D enstrophy spectra
  computed on circular shells for times matching the frames in figure
  \ref{UPO37_vort} and right, the evolution of energy in modes $\bm
  k_1=(1,0)$ (red solid) and $\bm k_2=(0,1)$ (black dashed). Symbols
  correspond to the frames in figure \ref{UPO37_vort}, $t=0$ star,
  $t=3.99$ triangle, $t=7.98$ circle and $t=15.01$ square.}
\end{center}
\end{figure}

A 1D spectrum of enstrophy (computed via circular shells in wavenumber
space) in figure \ref{fig:UPO37_k} (left) shows that the majority of
activity occurs for $k$ between the forcing scale and the largest
scale, however the flux to large $k$ near $t=7.98$ can clearly be
seen. Examining the transfer of energy amongst the large scales
reveals a simple process; the large-scale energy exchange is dominated
by the interactions between the two fundamental modes: $\bm k=(1,0)$ -
a `wave' mode - and $\bm k=(0,1)$ - a `zonal' (flow) mode.  In the
case of UPO 37 (figure \ref{fig:UPO37_k} right), the energy in the
wave mode $\bm k=(1,0)$ can account for up to $90\%$ of the total and
is inverse-correlated with the zonal mode $\bm k=(0,1)$ i.e. growth in
one is always associated with a decrease in the other. Physically, the large-scale
vortical motion (the culmination of the inverse cascade) corresponds
with both wave and zonal modes having comparable energy. The
destruction of the vortices (i.e. forward cascade) is associated with
the wave mode extracting energy from the zonal mode. The inverse
cascade, which occurs when the wave mode can no longer be sustained by
the weakened zonal mode, is signalled by the growth of the zonal flow
which reclaims it energy back from the wave mode to form large scale
vortices before the cycle repeats.

\begin{figure}
\begin{center}
\includegraphics[width=0.9\textwidth]{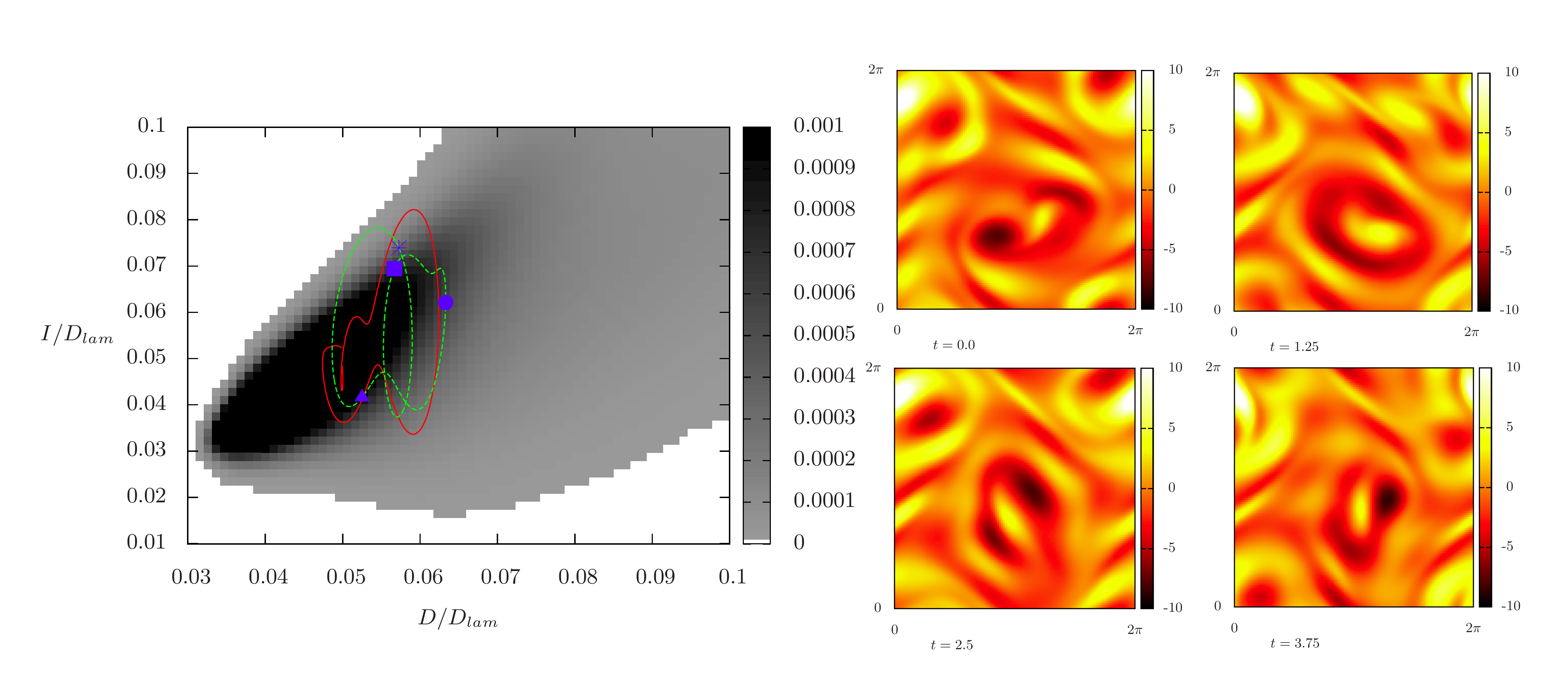}
\caption{\label{fig:DI_2} Left: projection of the flow onto the plane $I/D_{lam}$-vs-$D/D_{lam}$, greyscale showing the p.d.f. of the turbulent attractor, the green loop shows UPO 2 ( $T=5.096$ and $s=0.17$) and red open line the nearby segment of DNS which constituted the initial guess. Symbols correspond to the frames on the right: $t=0$ star, $t=1.25$ triangle, $t=2.5$ circle and $t=3.75$ square: vorticity plotted with shading as in figure \ref{UPO37_vort}. This example shows a relatively simple interaction between mid-scale vortices (vortex merging and filamentation) and consequently a compact periodic orbit in the sense of its extent relative to the underlying turbulent state (see left plot).}
\end{center}
\end{figure}

The close relationship between the wave and zonal  modes can be understood by making a simple general observation: in 2D, 2 Fourier modes
$\Omega_{\bm k_1}\exp(i {\bm k_1}.{\bm x})$ and $\Omega_{\bm k_2}\exp(i {\bm k_2}.{\bm x})$ cannot pass energy to a third Fourier mode $\Omega_{{\bm k_1}+{\bm k_2}} \exp(i({\bm k_1}+{\bm k_2}).{\bm x})$ {\em if} $|{\bm k_1}|= |{\bm k_2}|$.
To see this, consider the Fourier transform of the vorticity equation (\ref{NS}):
\beq
\dot{\Omega}_{{\bm k}} = 
\frac{1}{2}\sum\limits_{\bm k_1, \bm k_2 \in \mathbb{Z}} 
Z_{\bm k_1, \bm k_2}^{{\bm k}} \Omega_{\bm k_1}\Omega_{\bm k_2}
 -\frac{  |\bm k|^2 \Omega_{\bm k}}{Re} + \frac{n}{2} \delta_{ {\bm k} \pm n \hy} 
\eeq
where $\delta$ is the 2D Kronecker delta and $Z_{\bm k_1, \bm k_2}^{\bm k}$ is the interaction coefficient given by
\beq
Z_{\bm k_1, \bm k_2}^{\bm k} :=  \left(\frac{1}{ |\bm k_2|^2}-\frac{1}{ |\bm k_1|^2}\right)
({\bm k_1} {\bm \times} {\bm k_2}.\hz) \,\delta_{\bm k_1+ \bm k_2 - \bm k}.
\eeq
This clearly vanishes if $|{\bm k_1}|= |{\bm k_2}|$ which has two immediate consequences. The first is that Fourier modes in 2D cannot directly excite higher harmonics on their own (this observation is well-known in 3D for more general waves which have time-dependent wavevectors called Kelvin modes e.g. see \citep{CraikCriminale:1986} and references herein). The second is that energy placed entirely on a `ring' of wavenumbers defined by constant $|{\bm k}|$ will represent an exact nonlinear solution providing each mode satisfies its own {\em linear} equation (e.g \citep{Waleffe:1992}): for example, the 2D flow state
\beq
{\bm u}({\bm x},t)= \Re e \biggl\{\int^{2 \pi}_0 a(\theta) ( \sin \theta \hx -\cos \theta \hy) e^{ik(x\cos \theta+y\sin \theta)} d \theta \, \biggr\}
\eeq
is a steady solution of the 3D Euler's equation for any complex amplitude function $a(\theta)$ and $k$. 
In UPO 37 and the other UPOs, the energy of the flow is not so singularly distributed but is nevertheless largely confined to the inner wavenumber circle $|{\bm k}|=1$.
A severely truncated system consisting of only three modes $\bm k_1=(1,0)$, $\bm k_2=(0,1)$
and $\bm k_3=\bm k_1+\bm k_2$ (with complex conjugation denoted by $*$) where
\begin{align*}
\dot{\Omega}_{\bm k_1} &= Z_{-{\bm k_2}, {\bm k_3}}^{\bm k_1} \Omega_{\bm k_2}^* \Omega_{\bm k_3}, \\
\dot{\Omega}_{\bm k_2} &= Z_{-{\bm k_1}, {\bm k_3}}^{\bm k_2} \Omega_{\bm k_1}^* \Omega_{\bm k_3}, \\
\dot{\Omega}_{\bm k_3} &= Z_{\bm k_1, \bm k_2}^{\bm k_3} \Omega_{\bm k_1} \Omega_{\bm k_2}.
\end{align*}
illustrates the key point: $\dot{\Omega}_{\bm k_3}=0$ since $|{\bm k_1}|= |{\bm k_2}|$ but non-vanishing $\Omega_{\bm k_3}$ allows energy to be exchanged between $\Omega_{\bm k_1}$ and $\Omega_{\bm k_2}$. In the full system, of course, these modes receive further energetic contributions from other triads but these appear secondary (of much lower energy). The accumulation of energy in this innermost ring of wavenumbers
arises from the natural inverse cascade process. Once there, it becomes to some extent trapped as its principal route away - interactions between energetic modes - is blocked. This feature  is generic across the UPOs found except for a small group of UPOs possessing small fluctuations which exhibit modest interactions between localised vortices (e.g. UPO 2 which is the subject of  figure \ref{fig:DI_2}). The implication of this observation concerning the UPOs is that this confinement process in wavenumber space is also a central feature of 2D Kolmogorov turbulence.

\subsection{Cycle expansions}\label{sect:cycle}

We now examine how the extended set of relative periodic orbits (RPO: now
totalling 86 with the 5 found in \cite{Chandler:2013}) can be used to predict
properties of 2D turbulent Kolmogorov flow at $Re=60$. The basic idea
is to develop an appropriately weighted expansion across the set of
relative periodic orbits as follows
\beq
\Gamma_{prediction}^N:= \frac{\sum_{i=1}^N w_i \Gamma_i}{\sum_{i=1}^N w_i}
\label{expansion}
\eeq
where the property $\Gamma$ could be the mean dissipation rate, the
mean profile or a pdf, $\Gamma_i$ is the property value and $w_i$ the
weight for the $i$th RPO, and $N$ is a finite but large number (to be
discussed below). In periodic orbit theory, the weights cannot be
simply expressed but instead emerge from a recursive construction in
which the RPOs (`prime cycles') are arranged in progressively longer
sequences (`pseudo-cycles') which each contribute until a cut-off is
applied \citep{Cvitanovic13}. As in \cite{Chandler:2013}, the
dynamical-zeta-function periodic orbit averaging formula is used since
this only needs information about the (typically much smaller number
of) unstable Floquet multipliers rather than the whole spectrum.  
If $\Lambda_k^{(i)}:=e^{\Re
  e(\lambda^{(i)}_k) T_i}$ is the modulus of the $k$th Floquet
multiplier of the linearised (Jacobian) operator around the $i$th
recurrent flow of period $T_i$ ($\lambda_k^{(i)}$ the complex growth
rate), then the weight associated with the $i$th recurrent flow
depends on
\beq \Lambda_{i}:= \prod_{k, \Lambda_k^{(i)}>1}
\Lambda_k^{(i)}=\exp(\sum_{k \in \K_i} \Re e (\lambda_k^{(i)}) T_i).
\label{stabcoeff}
\eeq 
(see e.g. \cite{Gaspard97,Lan10}, \S 20 of \cite{Cvitanovic13})
where $\K_i$ is the set of $k$ such that $\Re e (\lambda_k^{(i)}) >0$. Numerically the complex growth rates are computed via Arnoldi iteration using the ARPACK library \cite{Lehoucq:1998jb}, converging extremal eigenvalues to a tolerance of $10^{-6}$.
For bounded flows (no trajectories escape - see \S 20.4.1 \cite{Cvitanovic13}), the zeta-function averaging formula takes the (relatively) simple
form
\beq
\Gamma_{prediction}^N:= \frac{\langle  \Gamma \rangle}{\langle T \rangle} 
\label{POT}
\eeq
where
\beq
\langle \Gamma \rangle:= \sum^{'}_{\pi} (-1)^{k+1} 
\frac{ \sum_{i=1}^k T_{p_i}\overline{\Gamma}_{p_i}}
{\Lambda_{p_1}\Lambda_{p_2} \cdots \Lambda_{p_k}}
\label{Gamma}
\eeq
and 
\beqa
\langle T \rangle &:=& \sum^{'}_{\pi} (-1)^{k+1} 
\frac{ \sum_{i=1}^k T_{p_i} }                           
{\Lambda_{p_1}\Lambda_{p_2} \cdots \Lambda_{p_k}}  \nonumber \\
&=& \sum_{i=1} \frac{T_i}{\Lambda_i}
-\sum_{i=1} \sum_{j=i+1} \frac{T_i+T_j}{\Lambda_i\Lambda_j}
+\sum_{i=1} \sum_{j=i+1}
\sum_{k=j+1}\frac{T_i+T_j+T_k}{\Lambda_i\Lambda_j \Lambda_k}-\ldots
\label{T}
\eeqa
(see \S20.4.1 \cite{Cvitanovic13}).  Here the subscript $p_i$
refers to the $p_i$\,th prime cycle (taken to be all the recurrent
flows identified), $\overline{\Gamma}_{p_i}$ is the temporal average of
the quantity $\Gamma$ over this cycle and $\sum^{'}_\pi$
represents a sum over all ($k=1,2,3,\ldots$) non-repeating, ordered
combinations of prime cycles making up a pseudo-cycle
(e.g. $\pi=(p_1,p_2,p_3,\ldots, p_k)$ represents a pseudo-cycle of
prime cycles $p_1$ to $p_k$ concatenated to create a total period of
$\sum_i^k T_{p_i}$).  Very roughly, the geometrical meaning of a
pseudo-cycle is that it is a sequence of shorter periodic orbits that
shadow a longer periodic orbit along the segments $p_1,p_2,\ldots,p_k$
with the relative minus signs ensuring shadowing cancellations. 

As in \citep{Chandler:2013}, a cut-off strategy is adopted based upon
stability \citep{Dahlqvist91, Dahlqvist94, Dettmann97}, \S 20.6
\cite{Cvitanovic13}) in which only pseudo-cycles with
\beq
\Lambda_{p_1} \Lambda_{p_2} \cdots \Lambda_{p_k} \leq \Lambda_{max}
\label{cutoff}
\eeq 
are included: this indirectly sets $N$. Usually $\lambda_{max}$
is set as $\max_i \Lambda_i$ across all the prime cycles
found. However, figure \ref{UPO_60}, which shows how $\Lambda_i$
increases across all the (suitably reordered) 86 relative periodic
flows known at $Re=60$, indicates that the 4 most unstable cycles are
outliers. As a result we took $\Lambda_{max}=2.98 \times 10^{7}$ which
corresponds to $\Lambda_i$ for the fifth most unstable cycle (shown as
a dashed black line in figure $\ref{UPO_60}$) for the calculations
shown. There is no discernable difference, however, to the results
based upon $\Lambda_{max}=\max_i \Lambda_i$. In fact, $\Lambda_{max}$
can be reduced to $O(10^{4})$ before any difference is noticed.

As a benchmark against which to assess the performance of periodic
orbit theory, we also considered a control protocol consisting of a
`democratic' equal weighting,
\beq
w_i:= 1 
\eeq
across all the relative periodic orbits (note in \cite{Chandler:2013}, 
this was protocol 3 and also included a very small number of equilibria and
travelling waves also found).  The key measures we look to predict are
pdfs of the total kinetic energy $E(t)$ and the total dissipation rate
$D(t)$ together with the profiles of the mean flow $\bar{u}(y)$ and
root-mean-square profiles of the fluctuation fields, $u_{rms}(y)$ and
$v_{rms}(y)$.

%
%

\begin{figure}
\begin{center}\setlength{\unitlength}{1cm}
\includegraphics[width=0.8\textwidth]{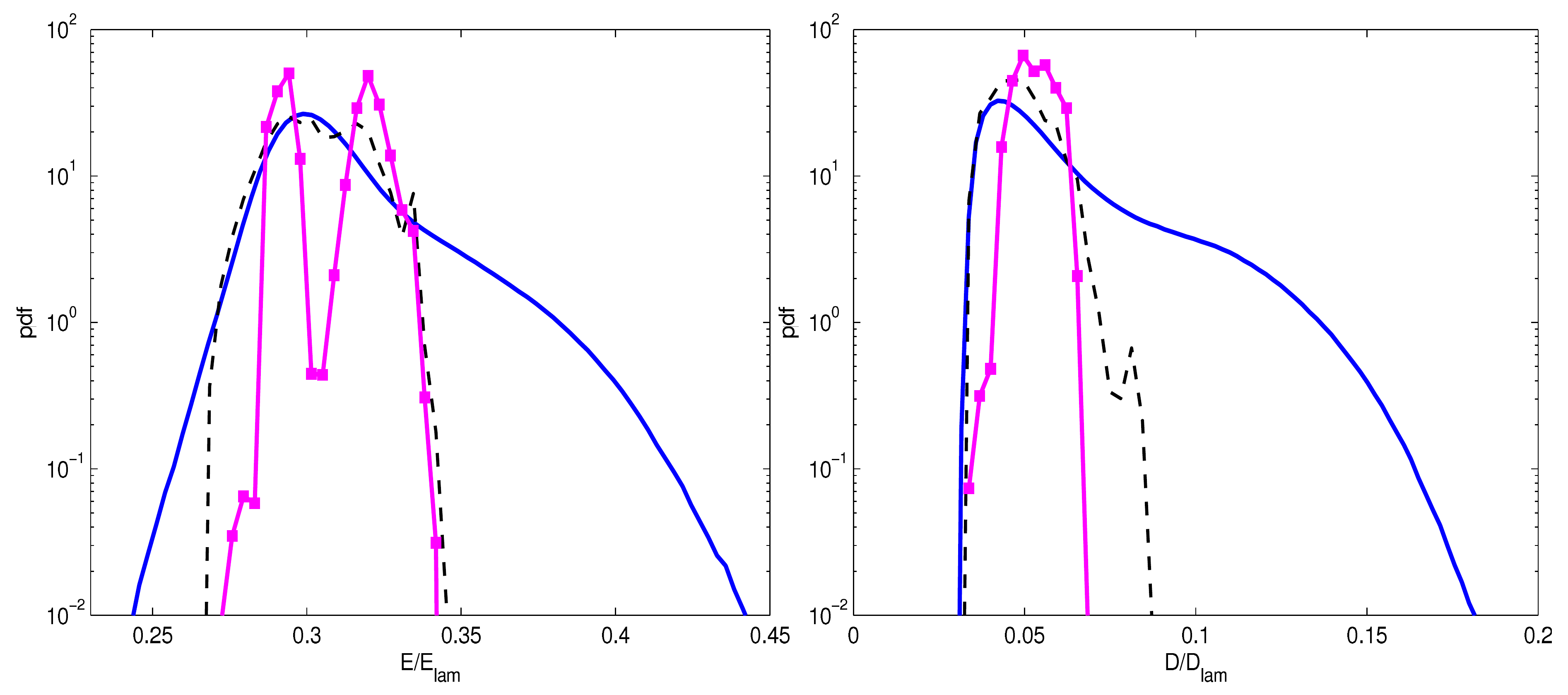}
\end{center}
\caption{The probability density functions for $E(t)/E_{lam}$ (left) and
  $D(t)/D_{lam}$ (right) from DNS (blue thick line) and predictions using
  periodic orbit theory (magenta, thick line with squares) and the control
  (black thick dashed) at $Re=60$. 60 bins were used to calculate the
  pdfs for the recurrent flows and 100 bins for the DNS due to its
  greater range. These choices gave the best balance of resolution
  with the data available. }
\label{pdf_60}
\end{figure}

\begin{figure}
\begin{center}\setlength{\unitlength}{1cm}                                          
\includegraphics[width=0.8\textwidth]{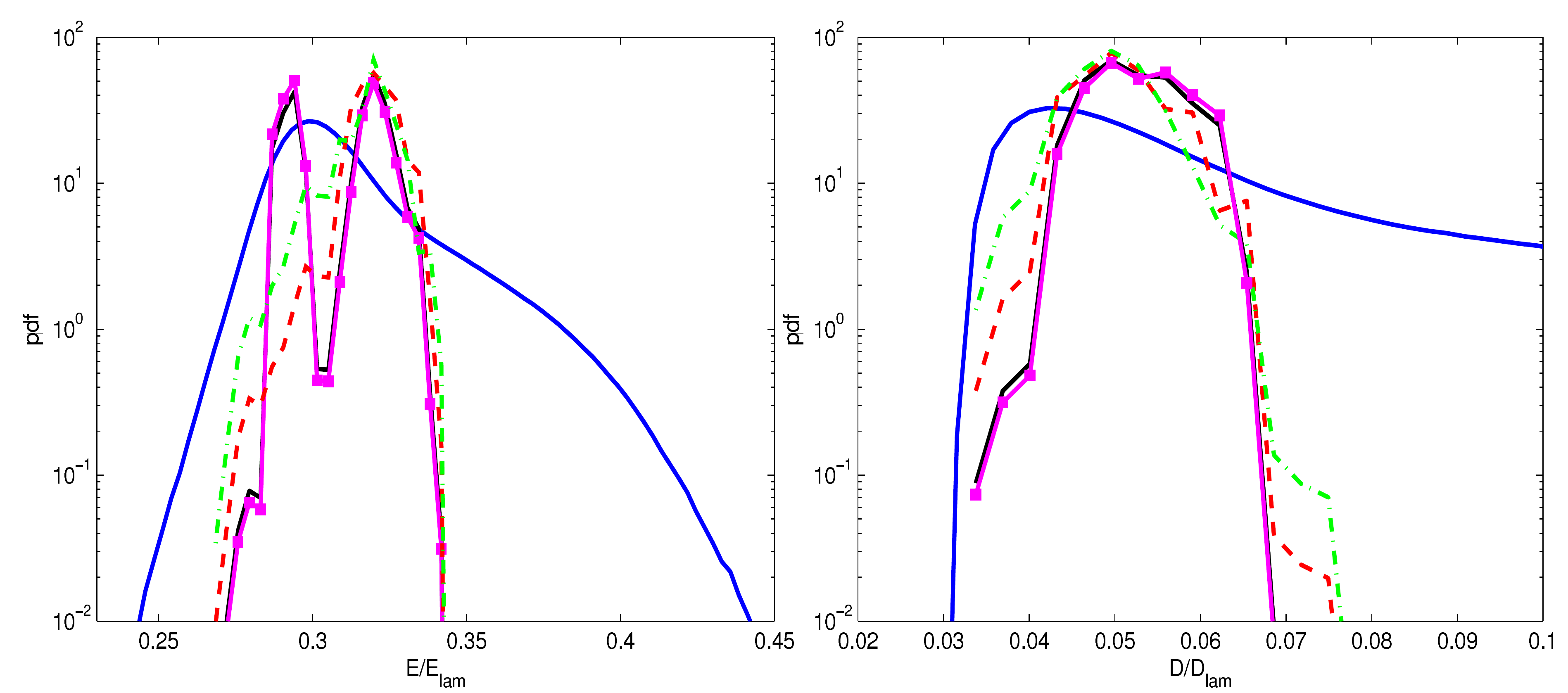}
\end{center}
\caption{The probability density functions for $E(t)/E_{lam}$ (left)
  and $D(t)/D_{lam}$ (right) from DNS (blue thick line) and the
  predictions from POT (magenta, solid line with squares), POT theory
  excluding pseudocycles (black solid line), POT with the first 4
  least unstable UPOs removed (red dashed line) and POT with the first
  11 least unstable UPOs removed (dash-dot green line).}
\label{pdf_60_play}
\end{figure}

The pdfs of $E(t)/E_{lam}$ and $D(t)/D_{lam}$ are shown in figure
\ref{pdf_60} along with the predictions using periodic orbit theory
(expression (\ref{POT}), magenta line with squares) and the control
protocol (dashed black line).  The control prediction is very good
across the maximum of the kinetic energy pdf but fails not
surprisingly to capture the lower shoulders ($E(t)/E_{lam}>0.36$ and
$E(t)/E_{lam}< 0.26$). The POT prediction is noticeably worse and
dominated by two peaks which indicate the presence of two different
groups of RPOs with separated energy levels. To confirm this
interpretation, the POT prediction was recalculated {\it excluding}
the first 4 least unstable RPOs (since they look like a separate group
stability-wise in figure \ref{UPO_60}) and the first 11 least unstable
RPOs (where again there is a discernible change in the stability of
the RPOs) with the results shown in figure \ref{pdf_60_play}. This
clearly shows that the first 4 least unstable RPOs are responsible for
the first peak in the pdf centred on $E/E_{lam} \approx 0.29$. Also
shown but difficult to distinguish from the full POT prediction is a
prediction based only on the prime cycles (black solid line).  This
indicates that including the pseudo-cycles has very little effect on
the scale of this plot.

The control prediction for the normalised dissipation rate pdf
$D(t)/D_{lam}$ is very good near the maximum but again fails to
capture the broad lower shoulder in the pdf at higher dissipation
rates. Similarly the POT prediction is not so effective,
overestimating the pdf peak and failing to capture the lower
dissipation behaviour of the dissipation rate as well as the
higher-dissipation shoulder. Interestingly, excluding the first group
of least unstable RPOs from the POT prediction improves matters at
least at low dissipation rates: see figure \ref{pdf_60_play}. Again,
there is little effect in excluding pseudo-cycles (see the overlapping
solid black line - just prime cycles included - and the magenta line
with squares - the full POT prediction in figure \ref{pdf_60_play}).

Given the poor representation of higher-dissipation DNS episodes, an
effort was made to revisit the data to particularly focus on these.
The original 11,934 guesses were filtered down to just 68 guesses
which had an average dissipation larger than the long time average of
the DNS calculation. These represent infrequent, high-dissipation
`bursting' events first noticed in \cite{Chandler:2013} and now
recognised in \cite{Lucas:2014} as intense interactions of spatially
localised `kink' and `anti-kink' regions (see \S \ref{rectangle}
below). Not surprisingly, these 68 guesses also had longer periods
(predominantly $T>40$) and none could be converged.  A revised
recurrent flow check was also tried in which large $k=|\bm k|$
wavenumbers were preferentially weighted to favour small scale/high
dissipation flows;
\begin{equation}
R(t,T) :=\min_{0\leq s \leq 2\pi/\alpha}\min_{m\in 0,1,2..n-1} 
\frac{\sum\limits_{k_x=0}^{32}\sum\limits_{k_y=-32}^{32}k\left | 
\Omega_{k_xk_y}(t)\mathrm{e}^{i(k_x\alpha s - 2k_ym\pi/n)}-
\Omega_{k_xk_y}(t-T) \right |^2}{\sum\limits_{k_x=0}^{32}\sum\limits_{k_y=-32}^{32}k \left | 
\Omega_{k_xk_y}(t) \right |^2} < R_{thres} \label{recur_highk}
\end{equation}
(note that the sums now need to retain more of the larger wavenumbers).
Taking a test run of $T=2\times 10^4$ and a threshold of $R_{thres}
=0.5$ produced 30 recurrent guesses with high dissipation although none, again,
could be converged.

%
%

%

%
%
\begin{figure}
\begin{center}\setlength{\unitlength}{1cm}
\includegraphics[width=0.8\textwidth]{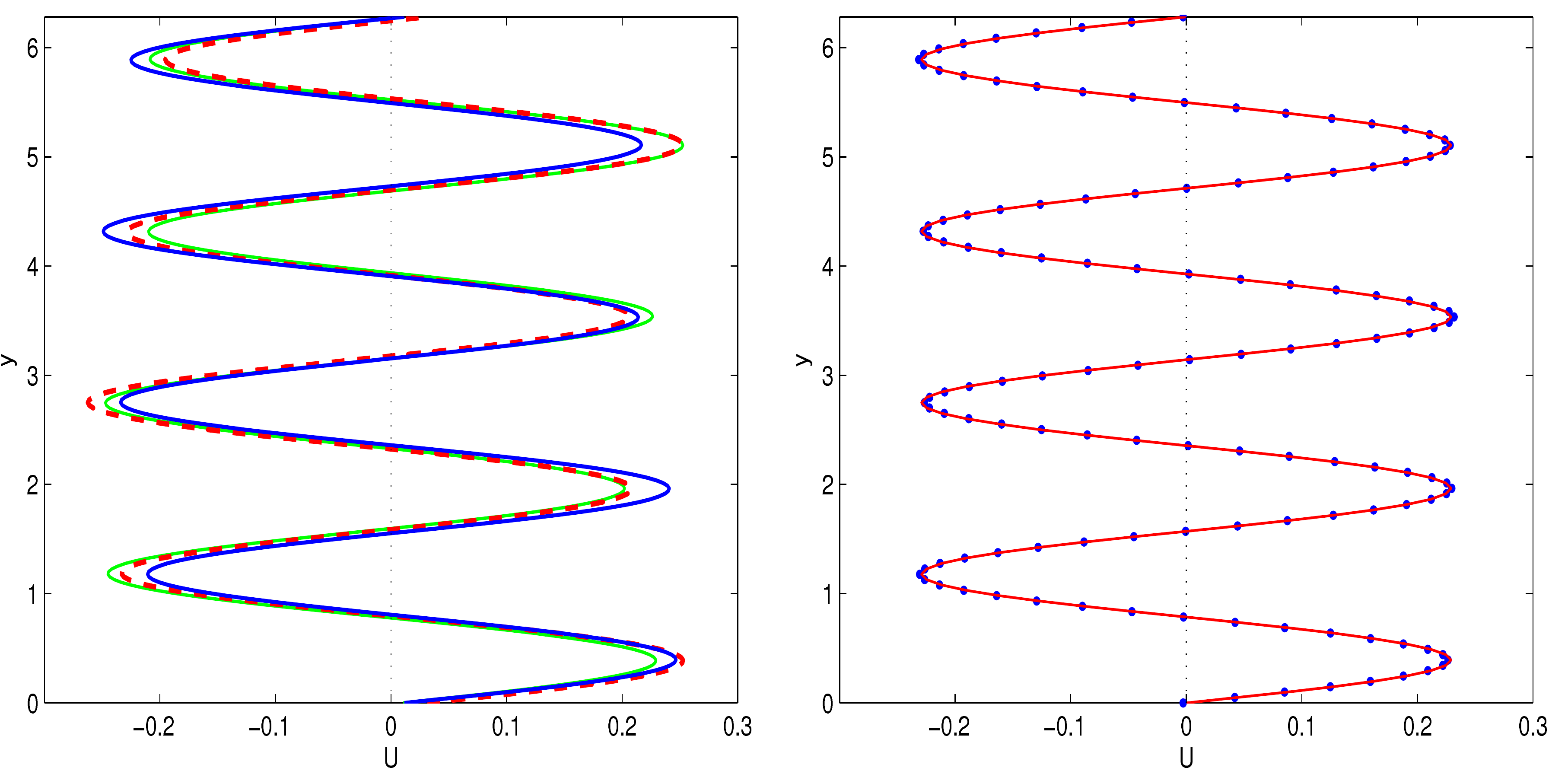}
\end{center}
\caption{Left: DNS mean flows from runs $e$, $f$ and $g$ (see Table 1
  of \cite{Chandler:2013}) at $Re=60$ showing a lack of convergence even after $10^5$
  time units (this is a reproduction of figure 22(b) of \cite{Chandler:2013}).  Right:
  DNS mean flow from $5 \times 10^{6}$ time units showing the expected
  symmetrisation: the DNS mean flow plotted across $[0,2\pi]$ using
  blue dots essentially coincides with the symmetrised DNS shown as a
  solid red line. }
\label{means}
\end{figure}

%
%
\begin{figure}
\begin{center}\setlength{\unitlength}{1cm}
\includegraphics[width=0.8\textwidth]{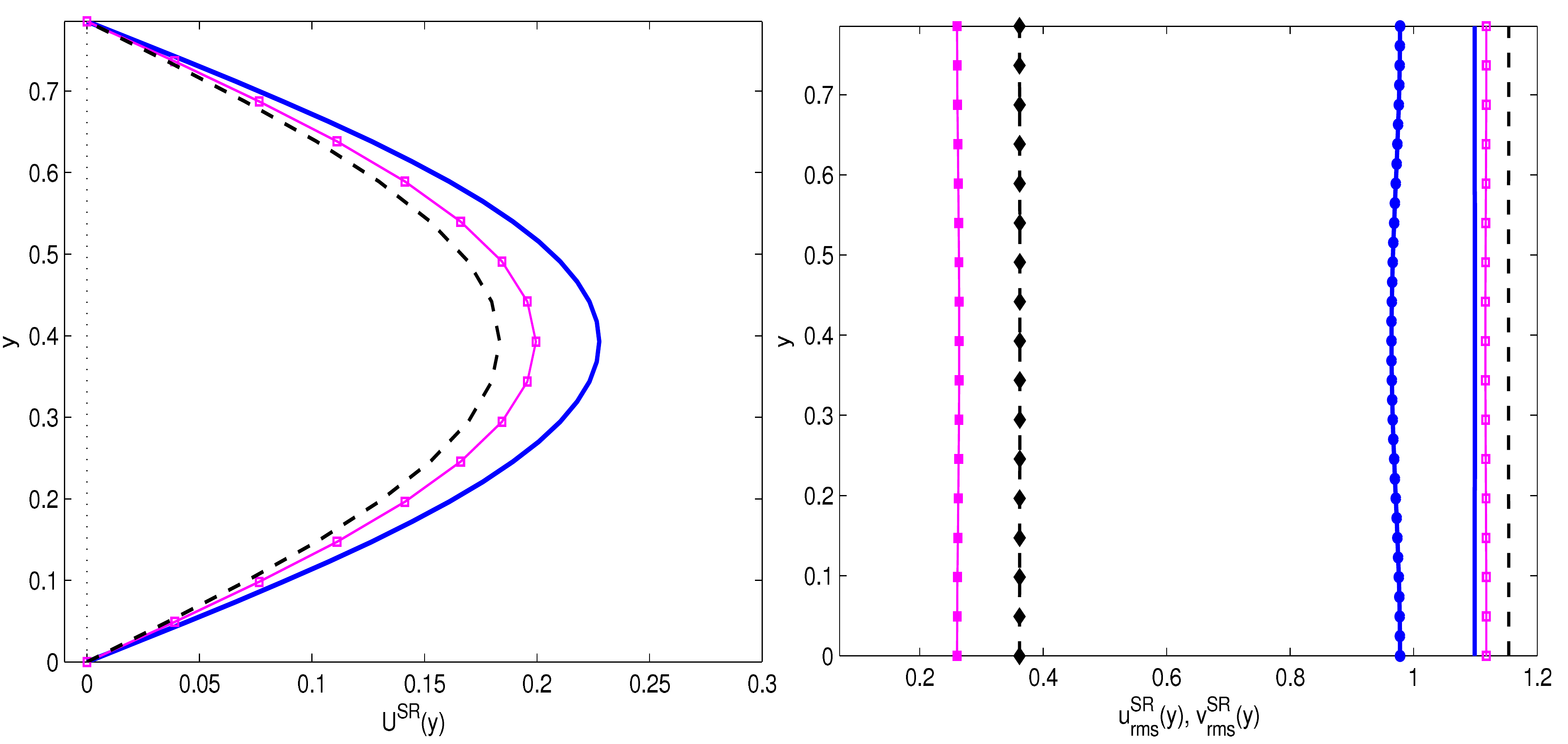}
\end{center}
\caption{Left: The symmetrised mean flow $U^{SR}(y)$ (DNS - solid
  blue), the POT prediction (magenta line with open squares) and
  control prediction (black dashed line) at $Re=60$. Right:
  symmetrised $u^{RS}_{rms}(y)$ (DNS - blue solid line with filled
  dots, POT prediction - magenta line with filled squares and control
  prediction - black dashed line with diamonds) and $v^{RS}_{rms}(y)$
  (DNS - blue solid line, POT prediction - magenta line with open
  squares and control prediction - dashed black line) over
  $[0,\pi/4]$. }
\label{pred_mean_60}
\end{figure}

Turning to the flow profiles, it was remarked in \cite{Chandler:2013}
that a run of $10^5$ time units at $Re=60$ was not long enough to see
a (long-time-averaged) mean flow sharing all the symmetries of the
system.  This assumed, by default, that the turbulent attractor is
unique and correspondingly that the mean should achieve this state of
full symmetry if the simulation was long enough. However, the
reduced-symmetry mean state found in \citep{Chandler:2013} could have
been one of many co-existing turbulent attractors selected by the
initial conditions.  It was therefore interesting to simulate the flow
for much longer - here $50 \times$ longer - to check the uniqueness of
the attractor. Figure \ref{means} (left) reproduces the 3 different
means generated in \citep{Chandler:2013} using 3 different initial
conditions each integrated over $1 \times 10^5$ time units which
clearly shows that they individually are not invariant under $\pi/2$
shifts in $y$ let alone under $\mathcal{S}$.  Figure
\ref{means} (right) shows the mean (plotted as blue dots) from the new
$5 \times 10^{6}$ data set. The red solid line (upon which the mean
blue dots basically sit) is the symmetrised version of the mean
defined as 
\beq 
U^{SR}(y):= \frac{1}{2n}\sum^{2n-1}_{m=0} {\cal
  S}^{-m}U^{R}({\cal S}^m y) \qquad {\rm with} \quad
U^{R}:=\frac{1}{2}[\,U(y)+{\cal R}^{-1}U({\cal R}y)\,]
\eeq (recall $n=4$) which satisfies all the symmetries of the flow
(this process picks out the following Fourier coefficients 
\beq
U^{SR}(y):= \sum_{m=0} a_m \sin(4(2m+1)y)
\label{symmetrised}
\eeq 
from the complete Fourier series of the mean flow
$U(y)$\,). The mean flow is essentially fully symmetric indicating
that the attractor is unique and that $5 \times 10^6$ should be
sufficiently long to generate worthwhile statistics. A similar
symmetrisation was also carried out for the root-mean-square profiles
of the fluctuation velocities to generate $u^{SR}_{rms}$ and
$v^{SR}_{rms}$.  Such symmetrised profiles need only be plotted over
$y \in [0,\pi/4]$ which is done in figure \ref{pred_mean_60} along
with the POT and control predictions. Both the POT and control
predictions perform well (with POT doing best) for $U^{SR}$ and
$v^{SR}_{rms}$ (the three lines to the far right of figure
\ref{pred_mean_60}) but are noticeably poor for $u^{SR}_{rms}$ which
has a value near 1 across the domain whereas the predictions never get
above 0.4 anywhere. The poor match in  $u^{SR}_{rms}$ could be related
to the lack of representation of high dissipation bursts in the recurrent flows
extracted. One such orbit was identified at $Re=40$ in \citep{Chandler:2013}
(named $R50$) and shown in figures 11 and 12 there. The large $D$ excursion
at approximately $t=8$ of this orbit can be seen to exhibit strong horizontal
velocity, suggestive that these events will contribute to the $u^{SR}_{rms}$ profile
heavily.

%
%

Some progress has been made from \citep{Chandler:2013} - we have
identified an order of magnitude more recurrent flows and the POT
predictions are improved (e.g. contrast figures \ref{pdf_60} and
26a from \cite{Chandler:2013}). However, periodic orbit theory is
still not outperforming a simple minded `democratic' averaging of the
RPO properties despite the nearly two orders of magnitude of RPOs
isolated.  The lumpiness of the kinetic energy pdf prediction in
particular highlights how a small number of weakly unstable RPOs can
dominate affairs due to the extreme sensitivity of the weighting of an
RPO to its stability (recall the Floquet multipliers are
exponentiated). Periodic orbit theory is clearly vulnerable to wide
disparities in stability across small sets of identified RPOs which
typically don't reflect the true distribution across the turbulent
attractor. This situation should, of course, be ameliorated by finding
even more RPOs - perhaps $O(1000)$ would emerge by increasing the
simulation time by another factor of 50 - but it also suggests some
better feel for phase space as a whole is needed to improve
matters. For example, are some of the longer period prime cycles
actually pseudo-cycles? (and therefore being counted incorrectly.) Are
all of the converged RPOs actually in the turbulent attractor or just
coincidentally converged from the DNS due to the vagaries of the
Newton algorithm in high dimensions? And where is the best place to
look to possibly `fill in' missing prime cycles (those of short period and
weak instability or those exhibiting the high $D$ bursting events) whose contribution is being overlooked?


\section{Results:  Long domain ($\alpha=1/4$)}\label{rectangle}

\begin{figure}
\begin{center}
\includegraphics[width=0.5\textwidth]{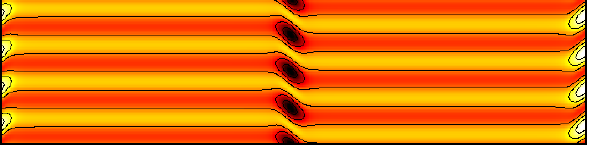}
\caption{\label{kink} Vorticity field, $\omega(x,y)$ at $Re=25$ for the steady solution branch which emerges from the initial bifurcation in the $\alpha=1/4$ domain (colour extrema are $\omega=-5$ black to $\omega=5$ white and 5 evenly spaced contours in $-4 \leq \omega \leq 4$). Two 1-dimensional regions are joined together by a `kink' (where the vorticity is a maximum i.e. to the side in the image) and an `antikink' (where the vorticity is most negative i.e. in  the middle here). }
\end{center}
\end{figure}

We now turn our attention to applying recurrent flow analysis to the
rich spatiotemporal behaviour found in \citep{Lucas:2014} when 2D
Kolmogorov flow is solved over a domain extended in the forcing
direction. The flow response mimicks the forcing at low forcing
amplitudes but, as in the square torus, beyond a critical value
develops a linear instability. This instability selects the longest
wavelength allowed with the flow displaying Cahn-Hilliard-type
(coarsening) dynamics for slightly higher $Re$ (see \cite{Lucas:2014}
and references herein). The unique attractor in this regime very
quickly (as $Re$ increases), develops two equal-length regions where
the flow is 1-dimensional joined together by localised 2D adjustment
regions - a `kink' and an `antikink': see figure \ref{kink}. Following this state
to higher $Re$, the kink and antikink pair suffer a sequence of
instabilities so that by $Re=40$, they are locally chaotic although
the 1D states which they connect remain steady. This chaos steadily
intensifies as $Re$ increases to $120$ and presents a good target for
recurrent flow analysis.  However, we start the discussion at lower
$Re$ by focussing on another state, $P1$, unearthed by
\citep{Lucas:2014} which exists in the window $11 \lesssim Re \lesssim
25$ where the flow possesses several attractors. The P1 orbit
is composed of a stationary kink-antikink pair flanking an inner
region where two oppositely signed vorticity patches (resembling a
kink-antikink pair in close proximity) interact to form an intertwined
oscillatory behaviour: see figure \ref{fig:snk_XT} for $(x,y)$ snapshots which resemble the flow. 
This inner region undergoes a period doubling
cascade into chaos as $Re$ increases leaving the flanking kink and
antikink pair still stationary.  A boundary crisis at $Re=24.1$ then
converts this attractor into a (localised) chaotic repellor.

\subsection{$P1$ at $Re=24$: chaotic attractor}

\begin{figure}
\begin{center}
\includegraphics[width=0.8\textwidth]{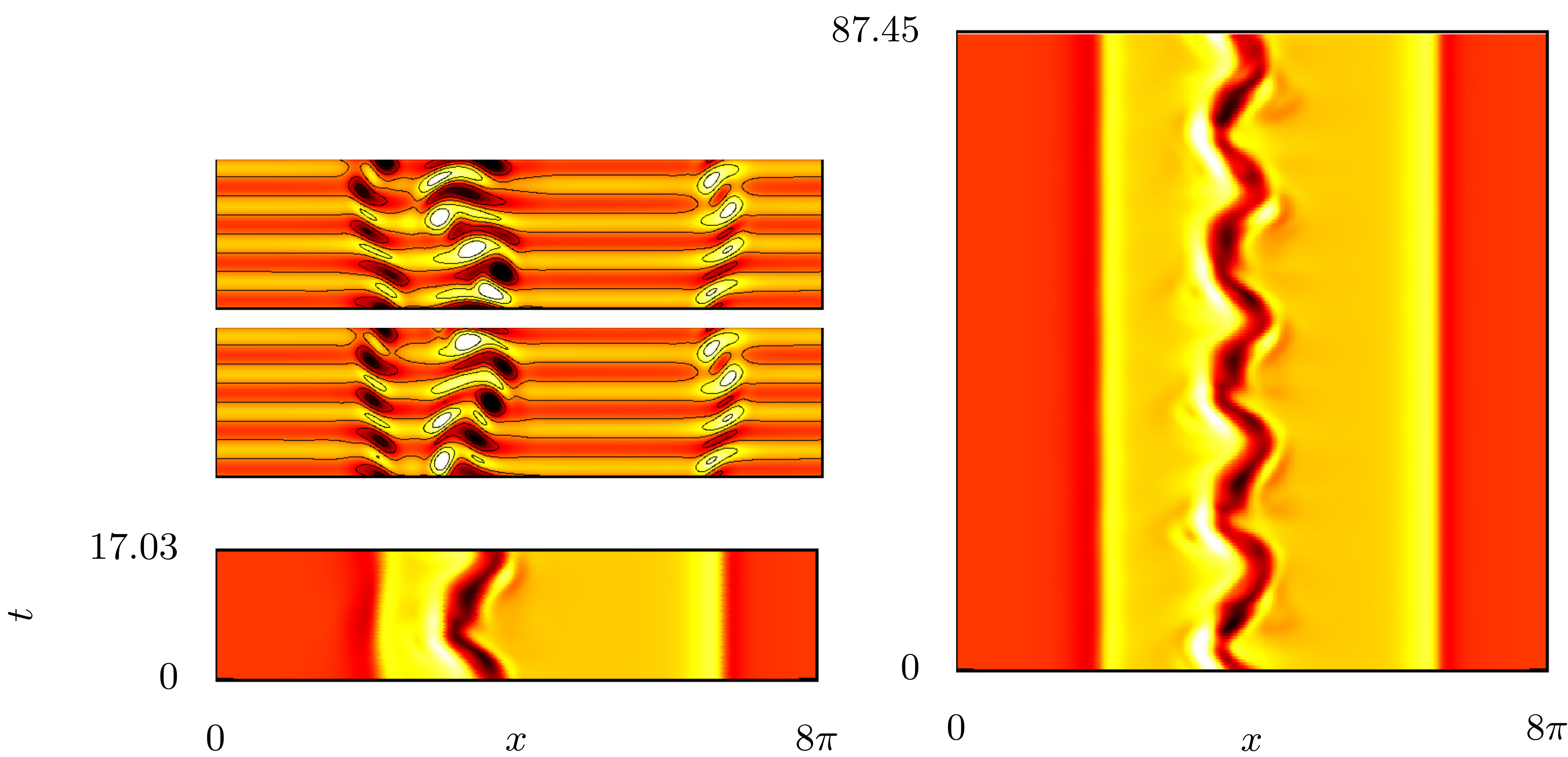}
\caption{\label{fig:snk_XT} Left: vorticity of the $Re=24$
  unstable orbit with $T=17.036$ in an $(x,t)$ plane with
  $y=21\pi/32$ (bottom) and $(x,y)$ frames at $t=0$ and $t=8.5$ (above). 
	Right: the orbit with $T=87.451$ in the $(x,y)$ plane (see Table \ref{snk_table}). 
	Colour extrema are $\omega = -5$ black, $\omega=5$
  white and $(x,t)$ plane plots have been scaled such that $t$ is
  approximately equal.}
\end{center}
\end{figure}

At $Re=24$, the P1 orbit is a spatially localised chaotic
attractor. Two DNS runs of duration $T=10^5$ were generated and the
reduced recurrence criterion (\ref{recur}) used to extract nearly
recurrent flows.  A threshold of $R_{thres}=0.3$ yielded 89 guesses
from which 2 unique periodic orbits were converged (3 convergences in
total). Lowering the threshold to $R_{thres}=0.2$ surprisingly yielded
an order of magnitude \emph{more} guesses, 935, from which another 2
periodic orbits were converged (23 convergences total). Table
\ref{snk_table} outlines the characteristics and convergence
frequencies of these 4 recurrent solutions (no steady states
were found). Figure \ref{fig:snk_XT} shows snapshots and space-time projections for two of the flows; 
the short period ($T=17.036$) flow is observed to exhibit a single internal oscillation, where the longer
orbit ($T=87.451$) has 5 such oscillations.

\begin{table}[htdp]
\begin{center}
\begin{tabular}{lcccccc}
$Re$ & $T$ & $s$ & $\sum\limits_{\substack{i \\ \Re e(\lambda_i)>0}} \Re e(\lambda_i)$ 
& $N$ &  freq. 1 & freq. 2 \\
\hline
24 & 17.036 &\quad 4.5E-4 &\quad 0.0624 &\quad 2 &\quad 1 &\quad 1 \\
24 &87.451 &\quad -1.9E-3 &\quad 0.0199&\quad 1 &\quad 2 &\quad 10 \\
24 &87.258 &\quad -7.5E-4 &\quad 0.0231 &\quad 2 &\quad 0 &\quad 10 \\
24 &87.634 &\quad -2.9E-4 &\quad 0.0147 &\quad 2 &\quad 0 &\quad 2 \\
\hline
24.5 & 18.405 & \quad  9.0E-4& \quad 0.2398 & \quad6 & \quad1 & \quad0\\
24.5 & 84.985 & \quad -1.3E-3 &\quad 0.0265   & \quad 2 & \quad1 & \quad$13$\\
24.5 & 16.986 & \quad 3.9E-4 & \quad 0.0807 & \quad 3 & \quad0 & \quad1 
\end{tabular}
\caption{\label{snk_table}Converged solutions from the $Re=24$
  P1-chaos attractor and $Re=24.5$ P1-chaotic repellor. Columns
  freq. 1 and 2 indicate the number of times the recurrent flow was
  discovered from the calculation with $R_{thres}=0.3$ and then $R_{thres}=0.2$
  respectively.}
\end{center}
\end{table}%

%
%

%
%

%
%
\begin{figure}
\begin{center}
\includegraphics[width=0.5\textwidth]{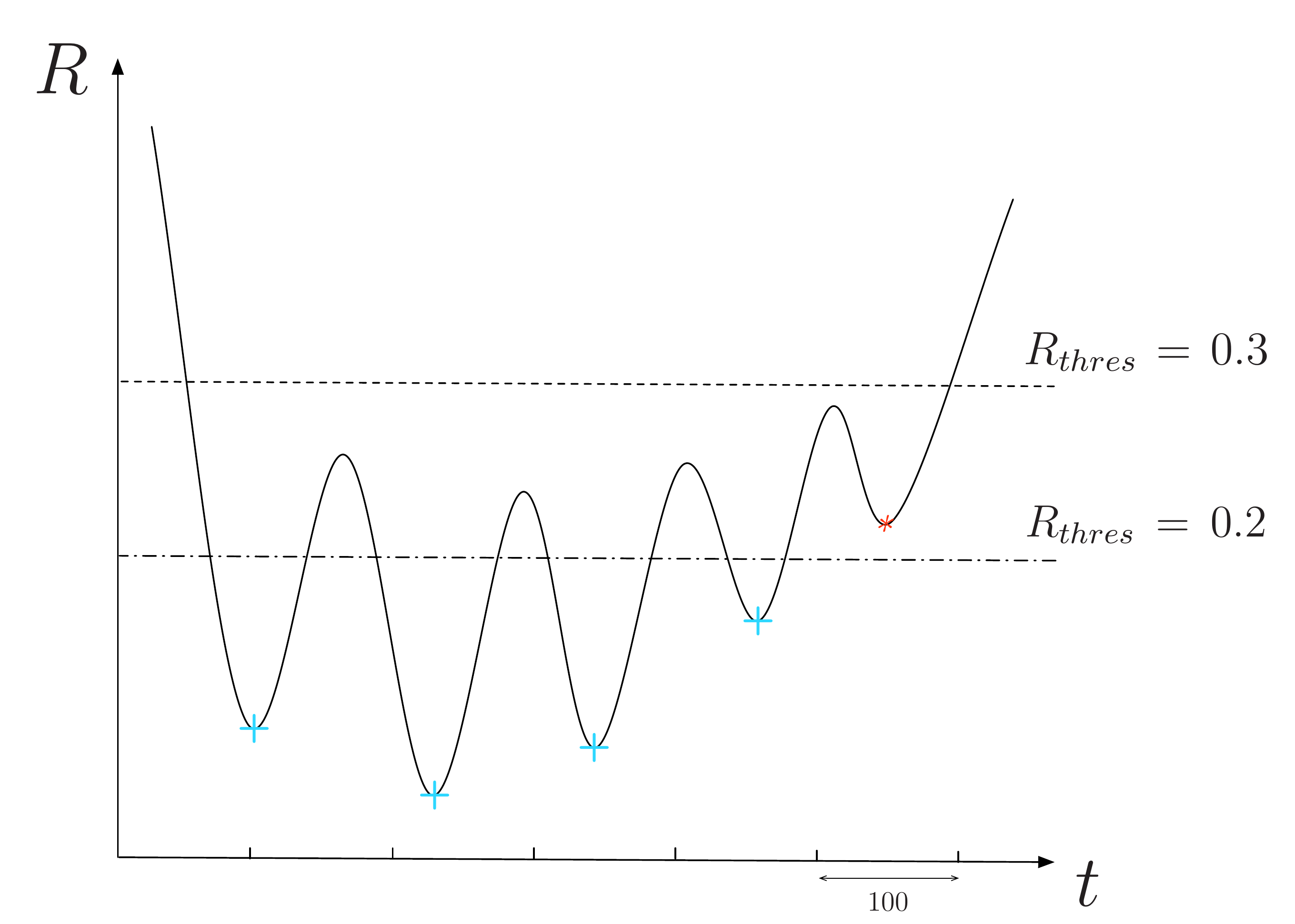}
\caption{\label{fig:resid}Schematic demonstrating how a smaller
  residual threshold can generate increased guesses. Red asterisk
  indicates guess generated by $R_{thres} = 0.3$ and cyan crosses
  guesses generated by $R_{thres}=0.2$.}
\end{center}
\end{figure} 
%
%
 \begin{figure}
\begin{center}
\includegraphics[width=0.8\textwidth]{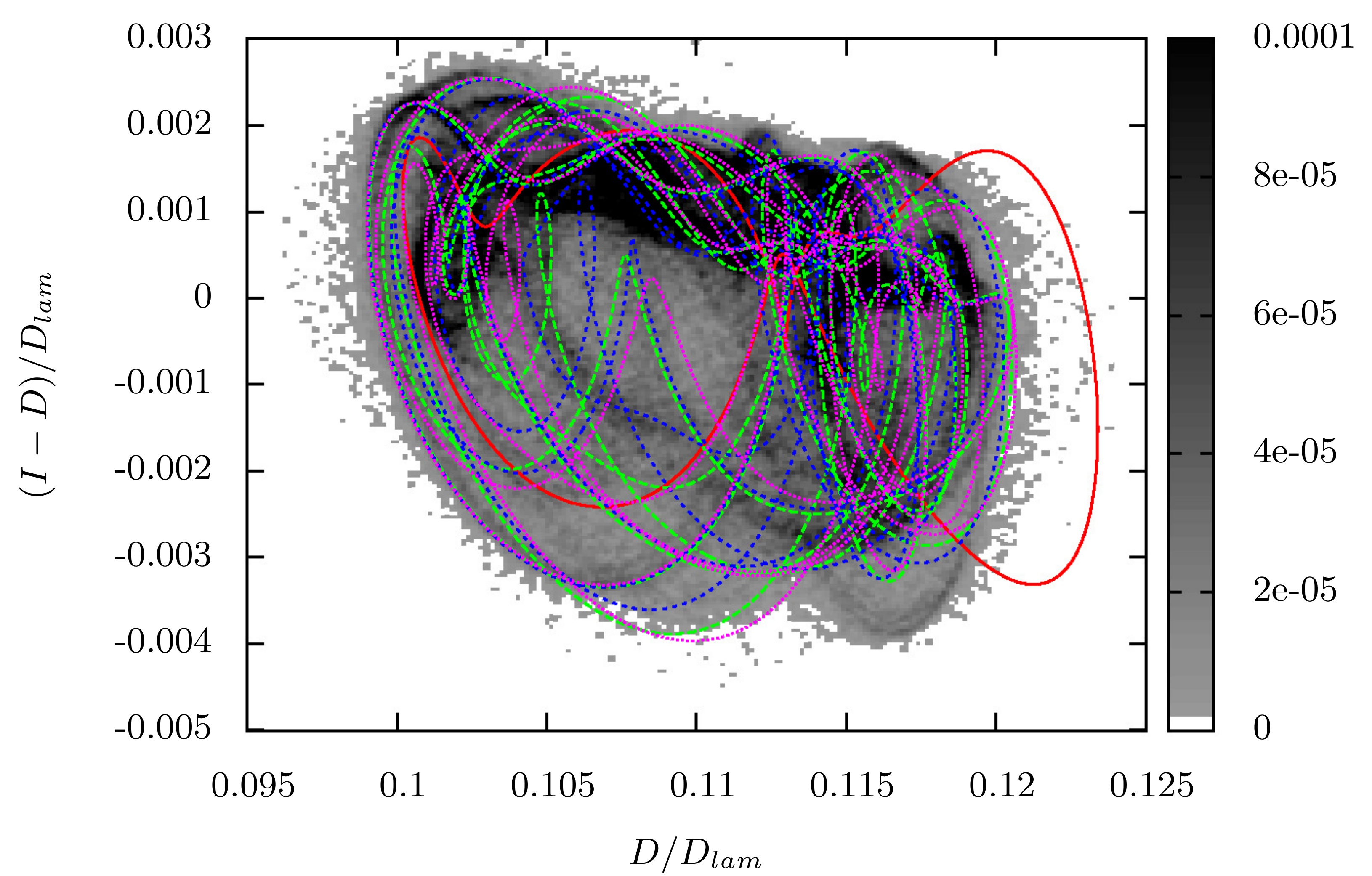}
\caption{\label{fig:snk_DI}$(I-D)/D_{lam}$ versus $D/D_{lam}$ for the
  converged orbits from the P1-chaos at $Re=24$. Grey shading shows
  the DNS pdf, the orbit with $T=17.036$ is red and the longer orbits
  with periods $\approx 87$ are drawn in green, blue and magenta.}
\end{center}
\end{figure}

The apparently counterintuitive inverse correlation between the threshold
level and the number of guesses found is actually due to the combination
of two effects.  The first is the fact that the recurrent flows at
$Re=24$ are only weakly unstable so that the flow spends a
comparatively long time in their vicinity potentially appearing as
multiple `visits' in one episode. The second is the necessarily finite
history window (chosen to be $[t-100,t]$) carried along by the
recurrence checking algorithm which is usually long enough 
compared to the periods of the extracted flow. When $R$ dips
below $R_{thres}$ indicating a nearly recurrent event, the algorithm
waits for $R$ to next exceed $R_{thres}$ before searching across the
retained history to store the velocity field at the smallest $R$.  If
there are a sequence of visits to an orbit such that $R$ repeatedly
dips below and above $R_{thres}$, multiple guesses will be stored.
If $R_{thres}$ is too high only the last visit will be stored
when only one visit fits into last 100 time units: see figure \ref{fig:resid}
for an example of this. This issue arises for weakly unstable
recurrent flows of long period as found here at $Re=24$.

\begin{figure}
\begin{center}\setlength{\unitlength}{1cm}                                          
\includegraphics[width=0.8\textwidth]{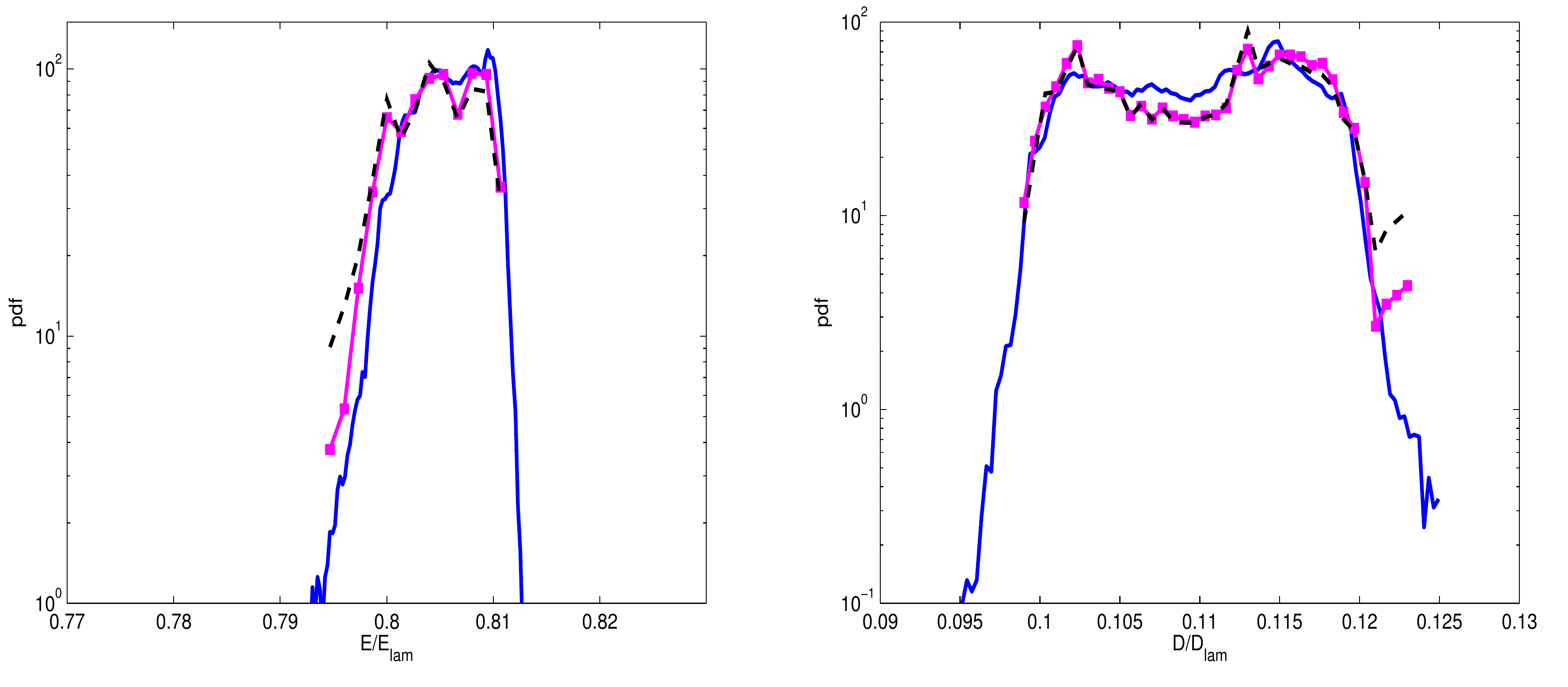}
\end{center}
\caption{\label{pdf_24}The probability density functions for
  $E(t)/E_{lam}$ (left) and $D(t)/D_{lam}$ (right) from DNS (blue
  thick line) and the predictions from POT (magenta, solid line with
  squares) and control (black dashed
  line). $Re=24$ P1 chaos and UPOs as in table \ref{snk_table}.}
\end{figure}

\begin{figure}
\begin{center}\setlength{\unitlength}{1cm}
\includegraphics[width=0.8\textwidth]{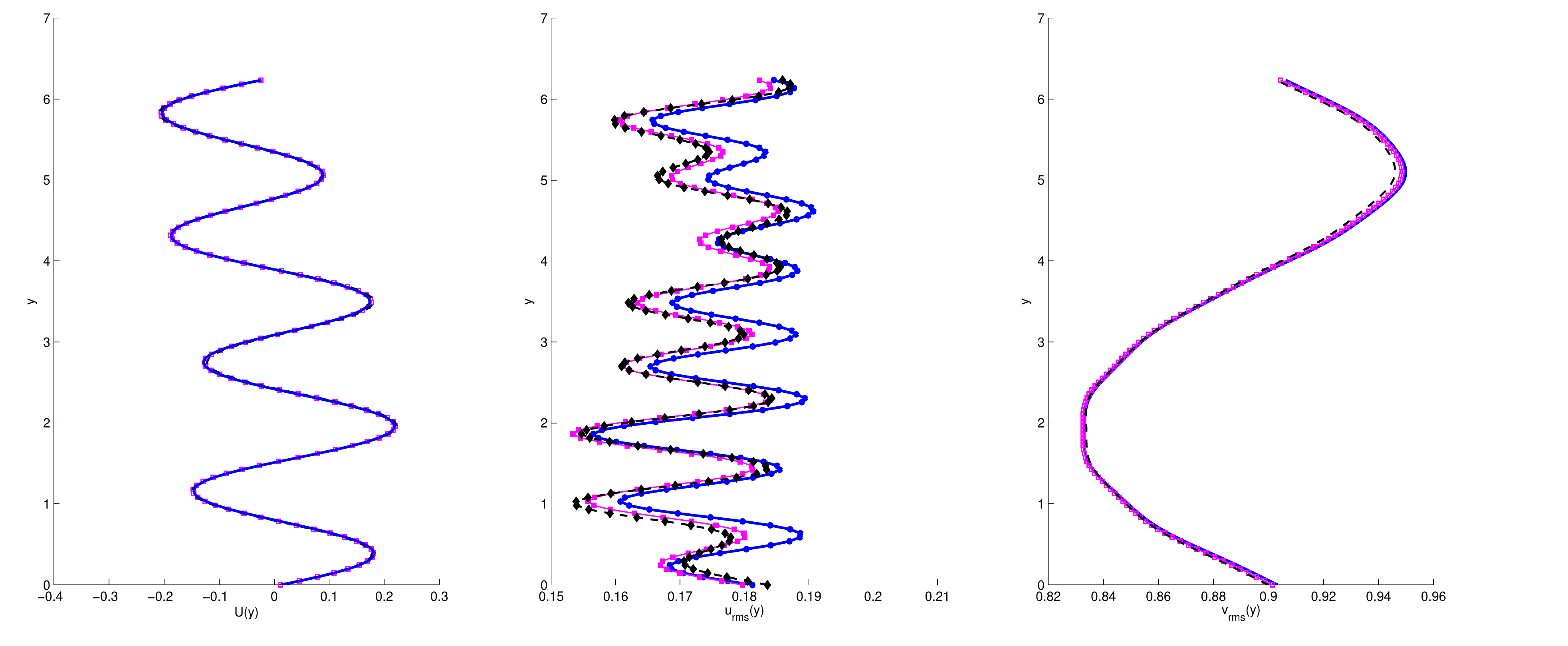}
\end{center}
\caption{\label{pred_mean_24} Left: The unsymmetrised mean flow $U(y)$
  (DNS - solid blue), the POT prediction (magenta line with open
  squares) and control prediction (black dashed line) for the P1-chaos
  at $Re=24$. Middle: unsymmetrised $u_{rms}(y)$ (DNS - blue solid line
  with filled dots, POT prediction - magenta line with filled squares
  and control prediction - black dashed line with diamonds).
  Right: $v_{rms}(y)$ (DNS - blue solid line, POT prediction - magenta line
  with open squares and control prediction - dashed black line) over
  $[0,2\pi]$.}
\end{figure}

\begin{figure}
\begin{center}\setlength{\unitlength}{1cm}
\includegraphics[width=0.8\textwidth]{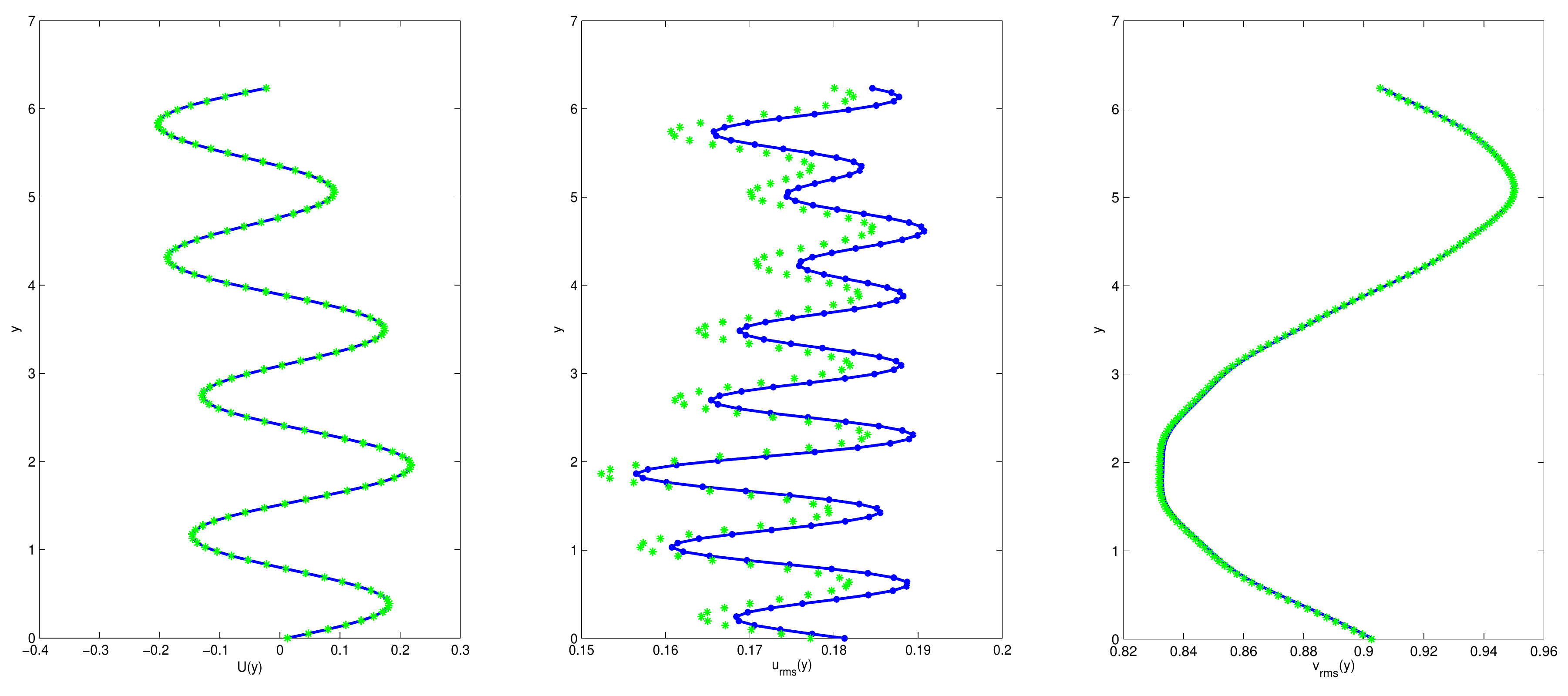}
\end{center}
\caption{\label{mean_UPO} Mean and r.m.s. velocities for the DNS
  (solid blue lines) and the recurrent flow with period $T=87.451$
  (green stars) with $Re=24.$ Left: The mean flow
  $U(y)$. Middle: $u_{rms}(y)$. Right: 
  $v_{rms}(y)$ over $[0,2\pi]$.}
\end{figure}

Figure \ref{fig:snk_DI} shows the probability density function for the
$P1$ chaos at $Re=24$ over the energy dissipation $D$ and the
difference between energy input and output, $I-D$, normalised by the
laminar dissipation (in the coordinates $(D,I-D)$ the pdf is centred
on the horizontal $I-D=0$ rather than the usual diagonal $I=D$ in the
$(I,D)$ plane). The four recurrent flows are also plotted and all look
fairly similar. The fact that the majority of recurrent episodes found
in the DNS had periods near to these orbits (figure \ref{fig:snk_gss}) suggests that periodic
orbit theory may work well. Using the cycle expansions of section
\ref{sect:cycle}, it was found that these few orbits reconstruct the
flow statistics impressively (see figures \ref{pdf_24} and
\ref{pred_mean_24}: note that the mean profiles are {\em not}
symmetrised and the control of democratic weighting does almost as
well). In fact, since each of the longer orbits spans the DNS pdf well
in figure \ref{fig:snk_DI}, the second recurrent flow with $T=87.451$
was chosen from Table \ref{snk_table} to compare its mean and rms
velocities with those of the DNS: see figure \ref{mean_UPO}.  The
profiles match remarkably well and recalls the earlier success of
\citep{Kawahara:2001ft} which appeared to demonstrate that a single
embedded recurrent flow was able to replicate the mean and rms
velocity profiles of  small-box turbulent plane Couette flow.

%
%

%
%

%
%

%
\begin{figure}
\begin{center}
\includegraphics[width=0.8\textwidth]{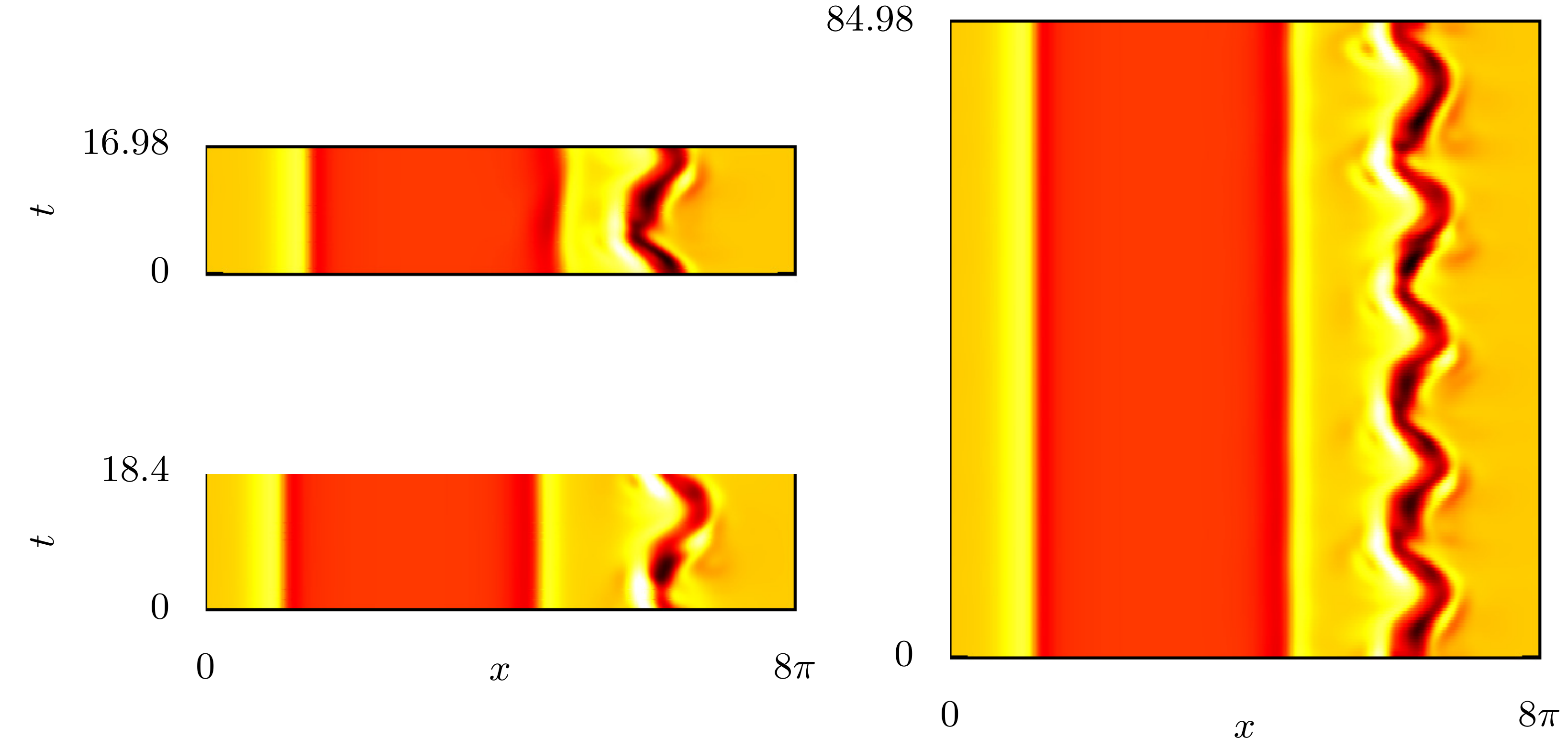}
\caption{\label{fig:snk_XTRe24pt5}Vorticity in an $(x,t)$ plane for
  $y=21\pi/32$ for the converged orbits from the P1-chaos at
  $Re=24.5$. Left shows the orbit with $T=18.405$ and right that with
  $T=84.985$. Colour extrema are $\omega = -5$ black, $\omega=5$
  white.}
\label{shown}
\end{center}
\end{figure}

%
%
\begin{figure}
\begin{center}
\includegraphics[width=0.8\textwidth]{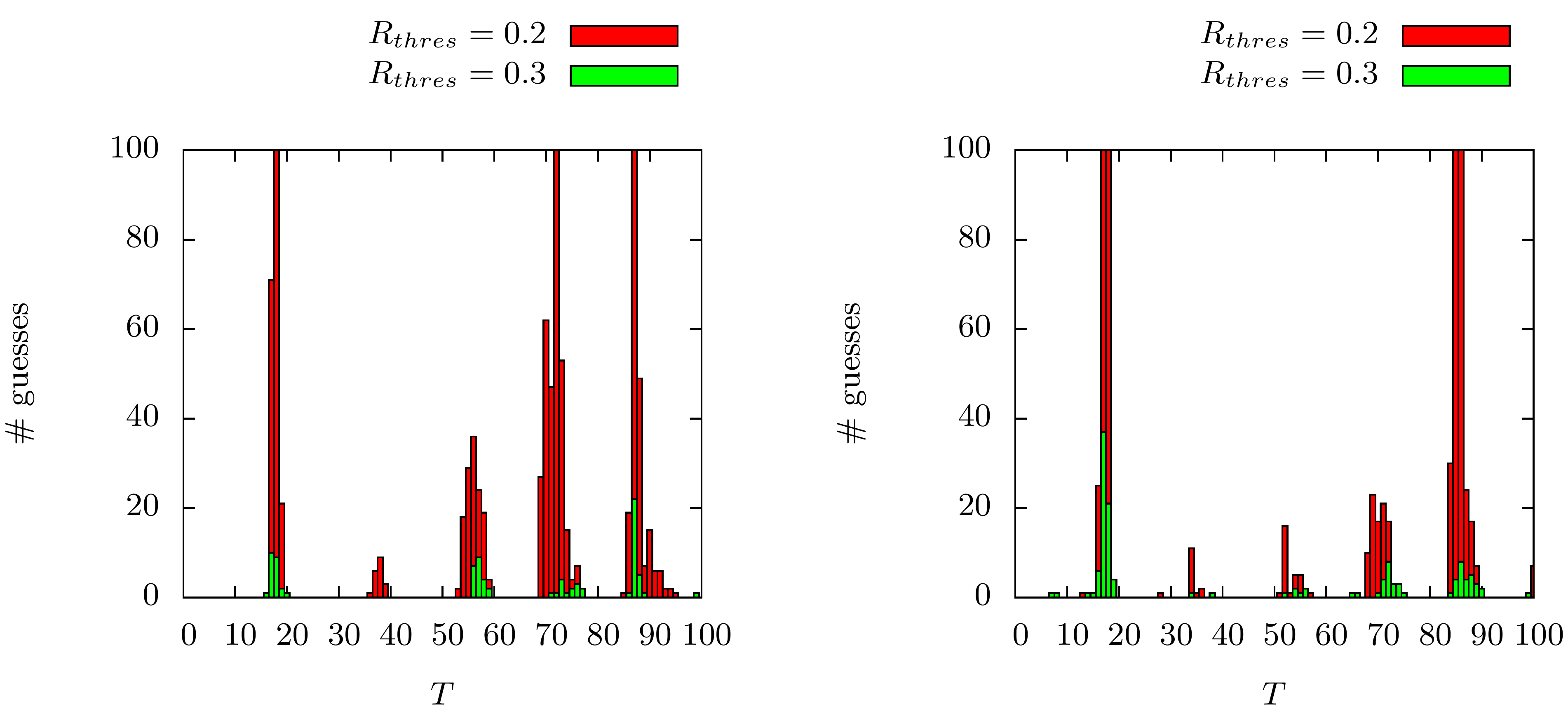}
\caption{\label{fig:snk_gss} Distribution of guesses with period for
  $R_{thres}=0.2$ and 0.3 for $Re=24$ (left) and $Re=24.5$
  (right). Notice the distinct recurrent periods found, as opposed to
  the broader band recurrences at $Re=60$, with $\alpha=1$.}
\end{center}
\end{figure}

\subsection{$P1$ at $Re=24.5$ chaotic transient}

\begin{figure}
\begin{center}
\includegraphics[width=0.8\textwidth]{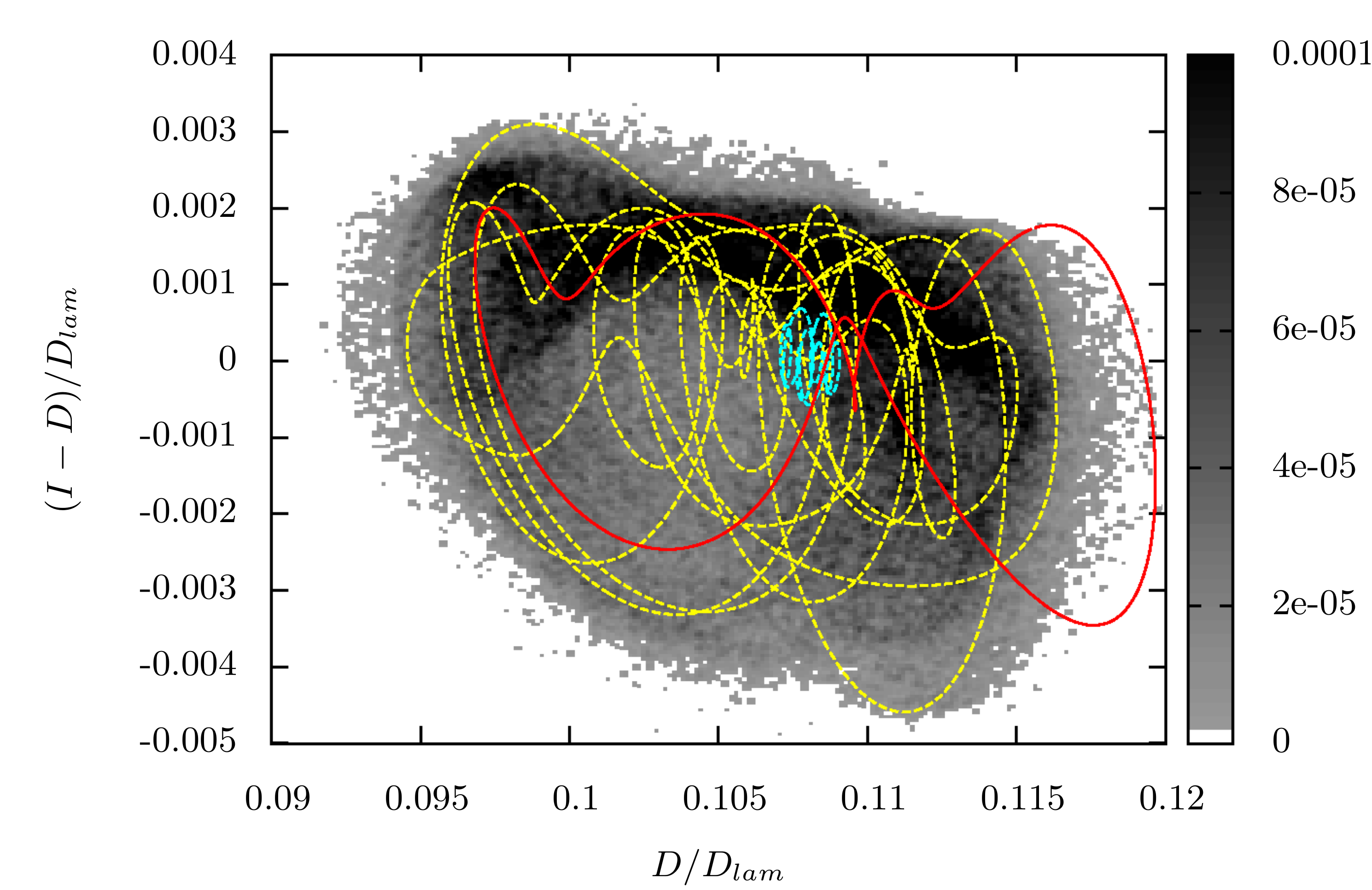}
\caption{\label{fig:Re24pt5_DI}$I-D$ versus $D$ for the converged
  orbits from the P1-chaos at $Re=24.5$. Grey colours show the DNS
  p.d.f., cyan shows the orbit with $T=18.405$, yellow that with
  $T=84.985$ and red the $T=16.98$ orbit as in figure
  \ref{fig:snk_DI}.}
\end{center}
\end{figure}

\begin{figure}
\begin{center}\setlength{\unitlength}{1cm}
\includegraphics[width=0.8\textwidth]{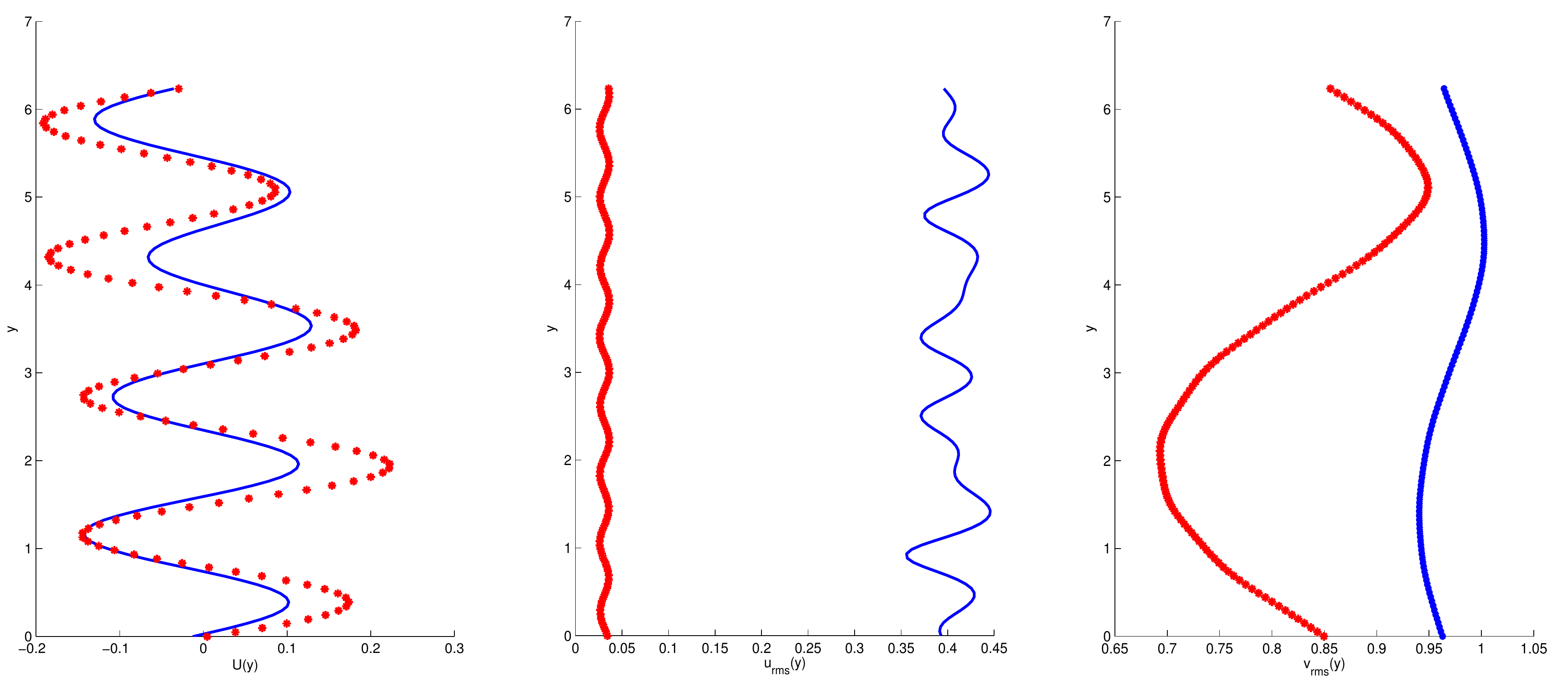}
\end{center}
\caption{\label{mean_24pt5} Mean and r.m.s. velocities for the DNS
  (solid blue lines) and the recurrent flow with period $T=84.985$
  (Red stars) at $Re=24.5$. Left: The unsymmetrised mean flow
  $U(y)$. Middle: unsymmetrised $u_{rms}(y)$. Right: unsymmetrised
  $v_{rms}(y)$ over $[0,\pi]$.}
\end{figure}

After a boundary crisis at $Re \approx 24.1$, the $P1$ chaotic
attractor becomes a chaotic saddle which was studied at $Re=24.5$
using recurrent flow analysis.  The lifetime of the transient chaos
here is $T\approx 4.1E4$ and thus still relatively long despite
apparently being close to the crisis. Setting $R_{thres} = 0.3$ yields
122 guesses from which only two recurrent flows were
converged. Lowering the threshold to $0.2$ gives 1842 recurrences from
which two orbits were converged; one new and the other a repeat of one
found at $R_{thres}=0.3$ (converged solutions are listed in table
\ref{snk_table} and shown in figure \ref{shown}). As with the
attracting chaos of the previous section, there are only a few
distinct periods within the set of recurrent guesses extracted from
the DNS (figure \ref{fig:snk_gss}).  However, this time, the three
orbits found have distinct characteristics. The $T=18.405$ orbit, when
viewed in the $(D,I-D)$ plane (figure \ref{fig:Re24pt5_DI}) is
visually more compact and does not span the attractor in the manner of
the rest of the unstable flows discovered at $Re=24$ or
$Re=24.5$. Branch continuing the orbits found at $Re=24.5$ down to
$Re=24$ revealed that the shorter period orbit, $T=16.986$ at
$Re=24.5$ is connected to the $T=17.036$ orbit at $Re=24$ but no other
connections were found.  The large period orbit, $T=84.985$, at
$Re=24.5$ appears in these figures similar to the longer orbits of the
sustained chaos at $Re=24$. However, in this case the mean profiles
(figure \ref{mean_24pt5}) do not match the DNS profiles.
Figure \ref{fig:snk_XTRe24pt5} shows the space-time projection for the three orbits.

%
%
%

%
%

%
%

\subsection{$P1$ at $Re=30$ chaotic transient}

\begin{figure}
\begin{center}
\includegraphics[width=0.8\textwidth]{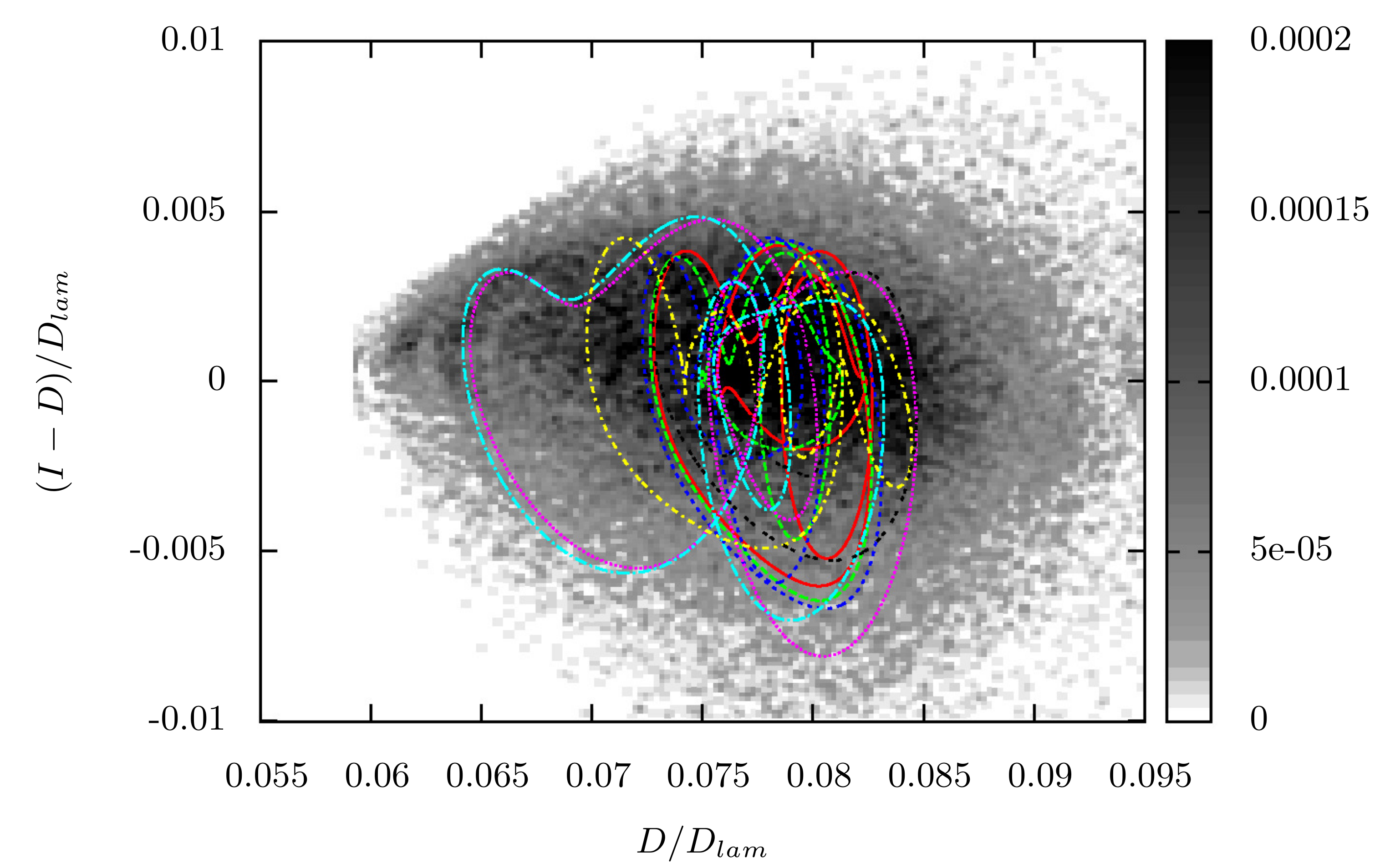}
\caption{\label{fig:Re30_DI} $I-D$ versus $D$ for the converged orbits
  from the P1-transient chaos at $Re=30$. Grey colours show the DNS p.d.f. with
  statistics accumulated over the 6 calculations.}
\end{center}

%
%
\end{figure}

\begin{figure}
\begin{center}\setlength{\unitlength}{1cm}                                          
\includegraphics[width=0.8\textwidth]{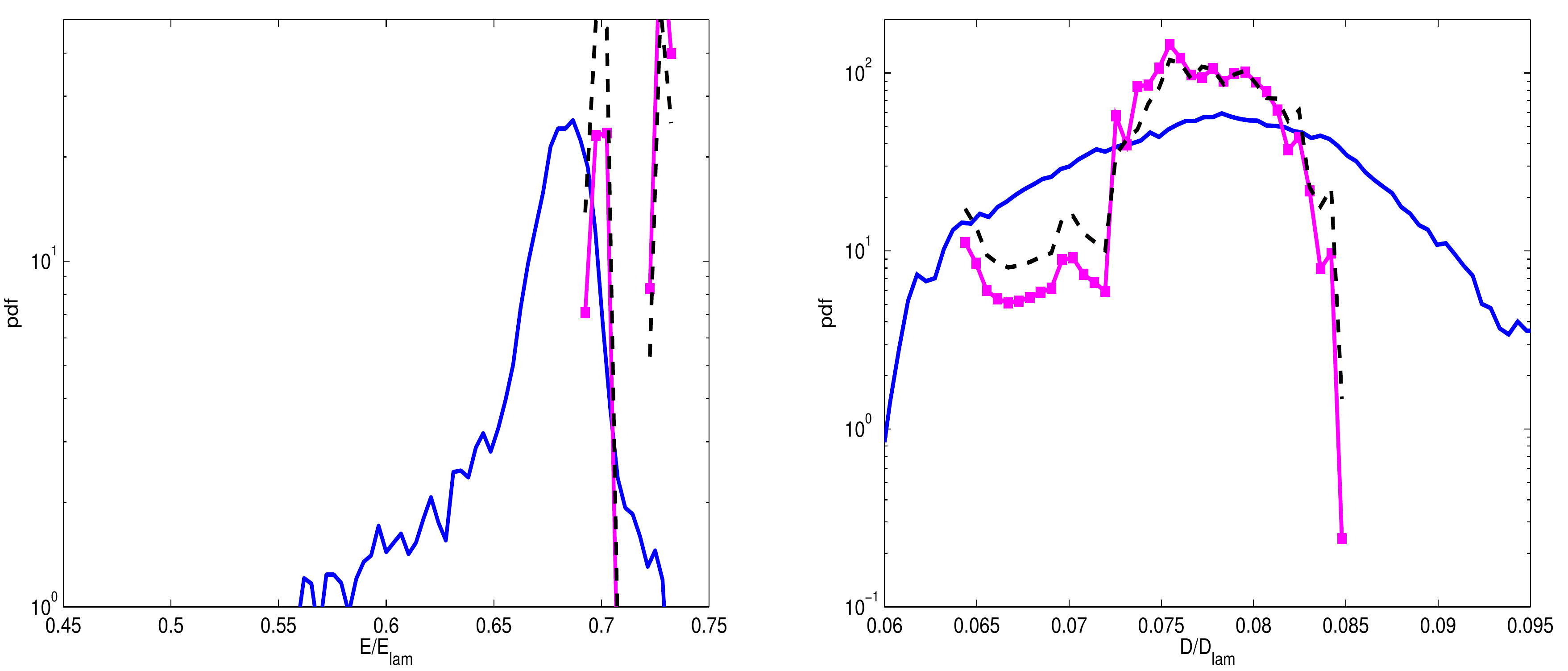}
\end{center}
\caption{The probability density functions for $E(t)/E_{lam}$ (left)
  and $D(t)/D_{lam}$ (right) from DNS at $Re=30$ of the $P1$-transient
  chaos (blue thick line) and the predictions from POT (magenta, solid
  line with squares) and the control (black dashed
  line).}
\label{pdf_30}
\end{figure}

\begin{figure}
\begin{center}\setlength{\unitlength}{1cm}
\includegraphics[width=0.8\textwidth]{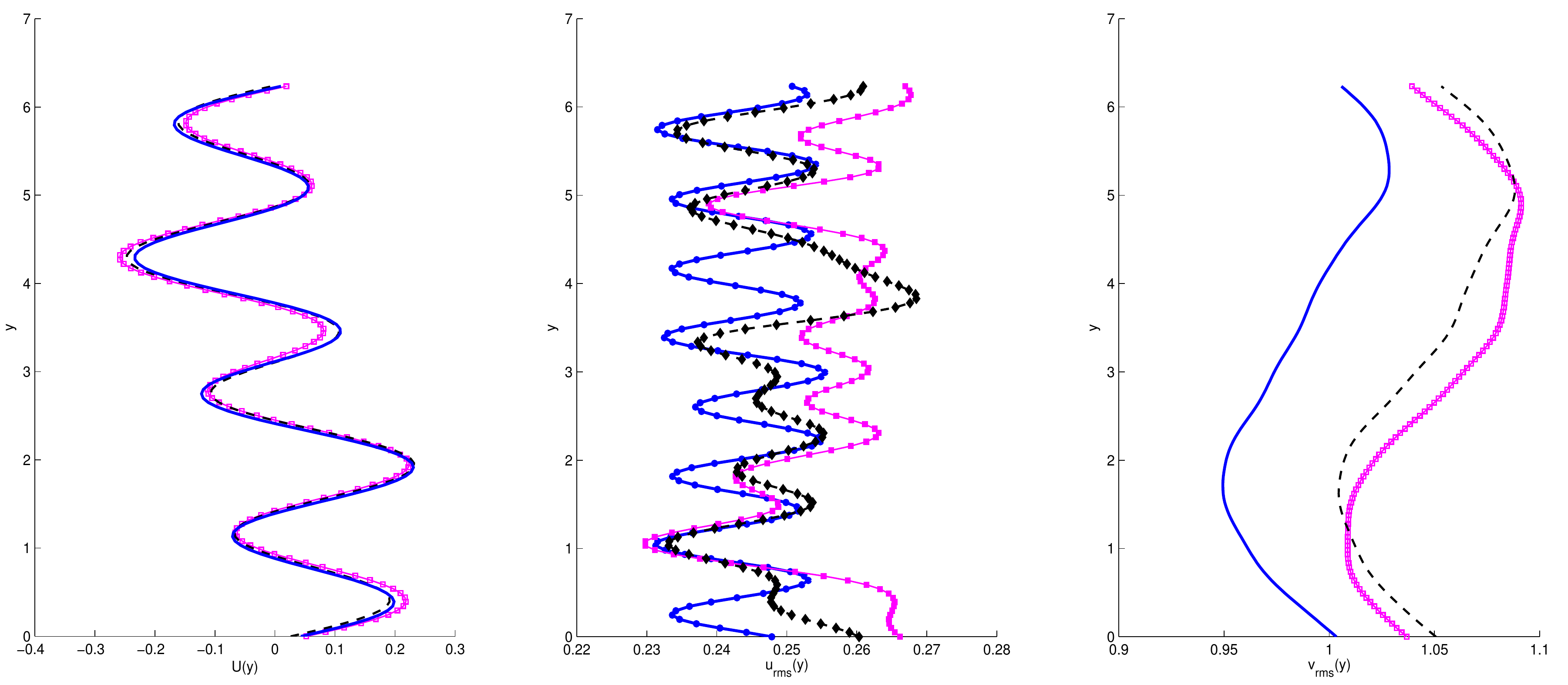}
\end{center}
\caption{Left: The mean flow $U(y)$ (DNS - solid blue line), the POT
  prediction (magenta line with open squares) and control prediction
  (black dashed line) at $Re=30$. Middle: unsymmetrised $u_{rms}(y)$
  (DNS - blue solid line, POT prediction - magenta
  line with filled squares and control prediction - black line
  with diamonds). Right: $v_{rms}(y)$ (DNS - blue solid line, POT
  prediction - magenta line with filled squares and control prediction -
  black line) over $[0,\pi/4]$. }
\label{pred_mean_30}
\end{figure}

At $Re=30$ the chaotic saddle has a much shorter mean lifetime and
therefore a suite of 6 DNS were performed with a randomly perturbed
(1\% with randomised phase) initial condition in each case in order to
build up a sufficiently long data sequence for recurrent flow
analysis. Across the 6 calculations, lifetimes ranged from $\approx
500$ time units up to $4500$ time units. The threshold $R_{thres}=0.3$
yielded 218 recurrence guesses across the concatenated data set. 
Of these, 7 (time-dependent) recurrent flows
were converged, each only once, as detailed in Table
\ref{snk30_table}. Figure \ref{fig:Re30_DI} shows these flows and the
probability density function for the chaotic repellor over the
$(D/D_{lam},(I-D)/D_{lam})$ plane. The coverage of the pdf by the
recurrent flows identified is much more limited than at lower $Re$
and, not surprisingly, the ability of these recurrent flows to
replicate the transient chaos is similarly reduced: see figure
\ref{pdf_30}. However, figure \ref{pred_mean_30} shows that the
velocity predictions are better with the mean profile reproduced best
(note the abscissa scales in figure \ref{pred_mean_30}).

\begin{table}[htdp]
\begin{center}
\begin{tabular}{ccccc}
$T$ & $s$ & $\sum \lambda$ & $N_\lambda$ & DNS\\
\hline
13.67 &\quad 2.3E-3 &\quad 0.216 &\quad 7 &\quad 1\\
17.24 &\quad 8.6E-3 &\quad 0.252 &\quad 4 &\quad 1\\
12.23 &\quad -7.0E-3 &\quad 0.359 &\quad 7 &\quad 2\\
12.11 &\quad -4.6E-3 &\quad 0.134 &\quad 5 &\quad 2\\
12.06 &\quad -7.4E-3 &\quad 0.0848 &\quad 4 &\quad 2\\
13.32 &\quad 1.1E-3 &\quad 0.212 &\quad 6 &\quad 3\\
13.07 &\quad 4.2E-4 &\quad 0.202 &\quad 4 &\quad 4\\
\end{tabular}
\caption{\label{snk30_table}Converged solutions from $Re=30$
  P1-chaotic repellor. DNS column indicates from which of the
  randomised initial conditions the unstable orbit was converged.}
\end{center}
\end{table}%

%
%

%
%

\subsection{$Re=40,70$ \& $120$ kink-antikink chaos}

The general trend from the above calculations is that while recurrent
flows can still be extracted as $Re$ increases, their predictive power
(at least in the numbers found here) reduces. This, of course, is
consistent with the usual increasing dimension of the chaos and
presumably with the increasing number of simple invariant sets
embedded in the chaos. In this section we consider yet higher $Re$
where the kink-antikink solutions, which are connected to the initial
bifurcation point in Kolmogorov flow, become chaotic.

At $Re=40$ with $R_{thres}=0.3$, we extracted 200 guesses from a
$T=10^5$ long DNS data set and converge 7 unique steady states (out of
81 convergences), all of which fall on the known solution branches
shown in figure 9 of \cite{Lucas:2014}. A test investigation at
$Re=70$ found similar results using a smaller temporal integration for
reasons of expediency (higher $Re$ requires a smaller timestep): from
$T=10^4$ with $R_{thres}=0.3$ we found 263 guesses which is an order of
magnitude more guesses in this more disordered regime from an order of
magnitude shorter data set. From these guesses, those with $R<0.1$ (54
guesses) were used to converge 2 distinct steady states (9
convergences) which again lie on known solution branches (figure 9 of
\cite{Lucas:2014}).

\begin{figure}
\begin{center}
\includegraphics[width=0.8\textwidth]{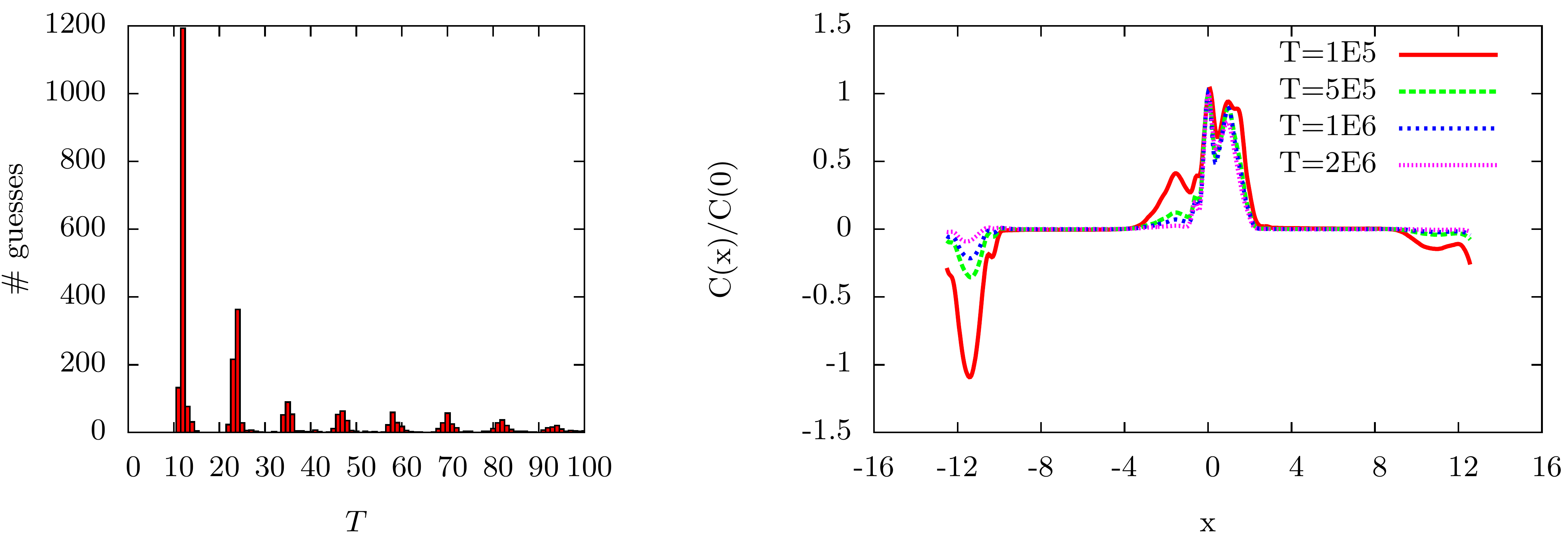}
\caption{\label{fig:Re120}Left plot shows the distribution of
  recurrence guesses for the $Re=120$ kink-antikink chaos with period,
  note the peak at around $T=11$ and subsequent multiples
  thereof. Right shows the two-point correlation of the $Re=120$ chaos
  plotted with distance from the peak vorticity.}
\end{center}
\end{figure}

Some effort was also made at $Re=120$: using a long $T=10^5$ data set
and $R_{thres}=0.3$ gave 2984 guesses. In contrast with the $Re=60$,
$\alpha=1$ calculations of section \ref{sect:60}, these were unevenly
distributed in period. Figure \ref{fig:Re120} shows that the majority
of recurrences found have a period close to 11 and the rest are
distributed around multiples of this value. This indicates
that there is a dominant recurrent flow (or set of similar flows)
which is expected to have a small number of unstable directions and is
therefore visited frequently in the chaotic attractor. For this
reason, and due to the higher computational burden at this Reynolds
number ($dt =0.002$), Newton-GMRES-hookstep efforts were concentrated
on guesses with $T\approx11$ but with no success: of 209 such guesses
attempted none converged. Several attempts were also made for higher
periods (e.g. $T\approx 22,$ 33 etc.) without success.

Given the spatial localisation of the chaos in the kink-antikink
solutions, one idea to improve the situation was to focus in on either the
kink or antikink chaos when extracting recurrent flow guesses. This
idea arose because, by $Re=120$, the chaotic kink and antikink look effectively uncorrelated
(except for the global constraint that the total vorticity is
time-invariant). To check this, figure \ref{fig:Re120} shows the
two-point correlation
\[ C(x-x_{max};x_{max},y_{0}) := \int\limits_0^T \hat{\omega}(x_{max},y_{0},t) \hat{\omega}(x,y_{0},t) \d t \]
(where $\hat{\omega}$ denotes the fluctuating part of the vorticity,
$x_{max}$ denotes the $x$ location of peak vorticity - the kink - and the choice
$y_{0}=\pi$ was made arbitrarily) normalised by the autocorrelation
$C(0;x_{max},y_0)$ for various time intervals.  The shrinking of the
  correlation near $x=-12$ (the antikink) with increasing $T$
  indicates that the kink and antikink regions are uncorrelated.
  Armed with this, a revised recurrent flow check was devised which
  only focussed on a subset (containing the kink) of the full
  domain.  This entailed working with the physical vorticity field
  $\omega$ in a region of width $\frac{15}{16}\pi$ (60 collocation
  points) centred on the supremum of $\omega$ across the full $y$
  range so that the residual function became
\begin{equation}
R(t,T) :=\min_{m\in 0,1,2..n-1} 
\frac{\sum\limits_{i_x=i_{max}-30}^{i_{max}+30}\sum\limits_{i_y=0}^{N_y}\left | 
\omega(i_x\frac{8\pi}{N_x},\, i_y\frac{2\pi}{N_y}- \frac{2m\pi}{n},\, t)-\omega(i_x\frac{8\pi}{N_x}, \, i_y\frac{2\pi}{N_y}, \, t-T)\right |^2}
{\sum\limits_{i_x=i_{max}-30}^{i_{max}+30}\sum\limits_{i_y=0}^{N_y}\left | 
\omega(i_x\frac{8\pi}{N_x}, \, i_y\frac{2\pi}{N_y}, \, t)\right |^2}
 < R_{thres} \label{recur_real}
\end{equation}
where $i_x$ and $i_y$ are the collocation points (respecting
periodicity at the boundaries) in the $x$ and $y$ directions
respectively and $i_{max}$ is the $x$ index of the supremum of
$\omega$. Since there was little drift of the chaotic attractor in $x$
(across the timescales of the DNS simulation), the shift $s$ was
assumed zero but allowed to vary from this value in the
Newton-GMRES-hookstep stage. Taking an integration length of $T=10^4$ and a threshold
$R_{thres}=0.2$ yielded over 200 such guesses, all of which again failed to converge.

In both choices of residual function, (\ref{recur}) and (\ref{recur_real}), almost identical behaviour of the Newton-GMRES-hookstep algorithm was found. Typically the residual saturated at values significantly lower ($10^{-3}$-$10^{-8}$) than the starting guess, but above the convergence criterion. This is suggestive that a limitation may exist at the Newton-GMRES stage rather than that the proximity of the individual guesses are poor.

%
%

\section{Discussion}

In this paper we have applied recurrent flow analysis to 2D Kolmogorov flow over a square and rectangular torus. In the former case, a much longer data set than previously \citep{Chandler:2013} has been generated in order to find more recurrent flows and thereby improve their predictive power as a whole. In one sense this has been successful - an order of magnitude more flows have been found - but in another, the set of flows  gathered still falls short of that necessary to produce a good prediction. Extrapolating, it would seem an $O(10^8)$ long data set should yield $O(1000)$ recurrent flows which should provide an even better prediction but exactly how much better is unclear. 

Over the rectangular torus, 2D Kolmogorov flow presents a number of interesting targets  for recurrent flow analysis since the flow exhibits localised chaos in various different ways. Recurrent flow analysis has been applied to these flows (for the first time to undeniably transient flows) over a range of $Re$ with varying degrees of success. At $Re=24$, the procedure has outperformed expectations with, in particular, just one recurrent flow proving a very good proxy for the chaos as in \citep{Kawahara:2001ft}.  
However, at larger $Re$ (here $\gtrsim 40$ for this extended domain), the picture quickly changes as it becomes more  difficult to find nearly recurrent flows and even harder to convert these guesses into exactly recurrent flows. Given this, our results indicate that the early success of \citep{Kawahara:2001ft} was more a feature of the weakly (low-$Re$) chaotic nature of their small-box plane Couette flow than a reflection of the true situation in stronger chaos/turbulence where the attractor dimension becomes large (e.g. \citep{Keefe1992}).

Algorithmically, efforts have also been made to streamline the process of identifying nearly-recurrent flows since this procedure is very costly compared to only performing the DNS. Clearly there are some savings to be made, at least working in spectral space. Some preliminary calculations have also been performed with a recurrent check criterion performed in physical space. This was motivated by the efficiencies that should be present in focussing on a subdomain of the flow which contains the active dynamics. Although nothing could subsequently be converged here, this is surely a promising area for future work, especially when the lengthscale of observed (localised) coherent structures differs significantly from that of the full flow domain (e.g. as $Re$ increases).  The one aspect of the recurrent flow analysis not investigated in any detail (except to adjust the various tolerances to help convergence)  is the Newton-GMRES-Hookstep algorithm. Its behaviour is well known to be complex and only really understood when it is started `very' close to an exact solution. Most importantly for the discussion at hand, it is possible for the algorithm  to converge to a simple invariant set in a very different part of phase space than where it started. This is clearly undesirable as an initial guess embedded deep in the turbulent attractor can lead to a converged recurrent flow outside of the attractor and therefore irrelevant for any prediction.  To avoid this scenario, a check could be instigated post-convergence to ensure the converged state is `close' to the initial guess but this was not done here.  

In terms of future directions, it should be clear that there are many challenges if the full programme of extracting recurrent flows directly from DNS data {\em and} then using them to predict turbulent statistics is to be pursued. Many will presumably be overcome, or at least reduced, by simply employing more computational power (e.g. the extraction of recurrent flows is readily automated). One outstanding question will, however, require more thought and perhaps a new idea/technique to emerge - how to weight the various contributions of the recurrent flows extracted.  It seems highly probable that more (global) insight from phase space is needed to inform this process than just the (local) stability information of the recurrent flow. Finally, if one simply wants to obtain a preliminary understanding of the flow, extracting and scrutinising the structure of extracted recurrent flows buried in the flow provides valuable information about what closed cycles of dynamical processes underpin the chaos/turbulence. This has been done here for the square torus flow and helped identify key dynamical features of the 2D turbulent flow. Future work will pursue this approach further in 3D turbulent Kolmogorov flow.

\noindent
{\em Acknowledgements.} We are grateful for numerous free days of GPU time on
`Emerald' (the e-Infrastructure South GPU supercomputer:
\url{http://www.einfrastructuresouth.ac.uk/cfi/emerald}) and the
support of EPSRC through grant EP/H010017/1.\\

\section*{Appendix A. \label{sect:crit}Recurrence criterion}

In section \ref{extraction} a considerable computational saving was
achieved by reducing the number of modes included in the recurrence
check. In this appendix we quantify such savings and the efficacy of
alternative strategies. A series of shorter ($T=2\times10^4$) DNS
calculations on the $Re=60$ attractor were performed each with a
modified recurrence check. To establish how efficient the
check was, attempts were also made to converge the guesses with the
Newton-GMRES-hookstep algorithm.  Applying this root-finding algroithm
is by far the most computationally expensive component of the
recurrent flow analysis so reducing the number of unconvergeable
guesses has a significant impact on overall efficiency.

In the first investigation, the number of modes included in the
recurrence check was varied starting with $8 \times 16$ (as in
expression (\ref{recur})), then using $32 \times 64$ (16 times as
large) and comparing the results with $42 \times 84$ (the full state
vector after de-aliasing) all with $R_{thres}=0.3$. The results are
outlined in table \ref{gss_table} which highlights the considerable
overhead of the recurrence checking: the $8 \times 16$ case `only'
doubles the DNS runtime whereas the full resolution case completely
swamps the basic DNS runtime.

In searching over all possible continuous shifts a discretisation of
$2\pi$ must be employed with 30 shifts adopted as default following
\cite{Chandler:2013}. Increasing this figure to 120 discrete shifts
resulted in more guesses and marginally more convergences but crucially
entailed more computations: see table \ref{gss_table}. Interestingly,
the more detailed recurrence check led to a different relative periodic
orbit being converged (R56 from \cite{Chandler:2013} with $T=16.326$)
instead of an orbit with $T=17.975$ in the default case of 30 shifts
from the {\em same} DNS data. This demonstrates clearly the sensitivity of the
Newton-GMRES-hookstep algorithm to the initial guess.

The expression \ref{recur} involving the full state vector is the
obvious choice since it is precisely this residual which the
Newton-GMRES attempts to converge to zero. However, it is not clear
whether this criterion will be the most efficient at picking out
guesses with higher likelihood of convergence; in other words
recurrences which are more likely to be close passes to underlying
unstable orbits.  An alternative guess criterion was tested based upon
 the streamfunction ($\psi = \Delta^{-1} \omega$) which
has the effect of producing a smoother field from which to define a
recurrence. Computationally this simply amounts to replacing $\Omega$
in (\ref{recur}) with $\Psi=\Omega/|\bm k |^2$, which effectively
weights the small wavenumbers more heavily. Table \ref{gss_table} shows the results
of this criterion (now with $R_{thres}=0.07$ to pick out a comparable
number of recurrences) and indicates a similar proportion of
convergences; 31 out of 59 guesses, compared to 21 out of 42 for the
default criterion. However, this new criterion was not only more
computationally expensive (more operations computing $\Psi$) but also
failed to converge any recurrences with $T>5.0$ (how these criteria
compare as $R_{thres}$ changes was not considered).

Finally, a two-fold strategy was tested to generate recurrent flow guesses. First,
during the DNS, a norm of difference in
absolute value of the components:
\begin{equation}
R(t,T) :=\frac{\sum\limits_{k_x=0}^{8}\sum\limits_{k_y=-8}^{8}\left | 
\Omega_{k_xk_y}(t)\right|^2-\left| \Omega_{k_xk_y}(t-T) 
\right |^2}{\sum\limits_{k_x=0}^{8}\sum\limits_{k_y=-8}^{8}\left | 
\Omega_{k_xk_y}(t)\right |^2} \label{recur_abs}
\end{equation}
was used as a necessary but not sufficient recurrence condition. This
avoids the necessity to sweep over shifts, however, having generated a
`loose' guess in this fashion, there is a need to post-process to compute the
necessary shifts before attempting convergence. For this reason, the
CPU time quoted in table \ref{gss_table} includes the post-processing
routine. This is a first attempt at a two-step strategy; we generated
a large set of `quasi'-guesses initially (2755) from which to process
more accurate guesses. Due to the size of this set the post-processing
took a considerable length of time, (257 CPU minutes) much longer in
fact than the DNS. In retrospect, a smaller set of quasi-guesses and
outputting both start and end points of the recurrence, would make
this process more efficient. In any case, from 25 secondary guesses we
converge only 5 recurrent flows. One has $T>5$ and
remarkably this is yet another new unique orbit which we have not
found in any previous searches using the same DNS data.

\begin{table}[htdp]
\begin{center}
\begin{tabular}{ccccc}
criterion & \# guesses & CPU time & \quad factor &\quad convergences \\
\hline
DNS          &  - & 27.7  & -     & -\\
$8\times16$  & 42 & 55.5  & 1     & 21\\
$32\times64$ & 35 & 456.2 & 8.2   & 16\\
$42\times84$ & 35 & 766.4 & 13.8  & 16\\
$N_s=120$    & 69 & 138.5 &  2.5  & 25\\
$\psi$       & 59 & 87.9  &  1.6  & 31\\
$|\Omega|$ & 25 & 28.1 (+257) & 5.1 &5\\
\end{tabular}
\caption{\label{gss_table}Table outlining successfulness of various
  recurrence checks and variants. First entry has DNS CPU time without
  recurrence checking for comparison, the following three show the
  standard check with various sizes (N.B. $42\times84$ represents the
  full state vector), then standard check with higher fidelity $s$
  shift search, using the streamfunction $\psi$ and final check is a
  residual based on absolute values of components. Using $8\times16$
  rather than the full $42\times84$ gives a 13.8 times speed up.}
\end{center}
\end{table}%

\section*{Appendix B}

In this appendix, details of all the recurrent flows found in section \ref{sect:60} are given 
in a table.

\begin{table}[htdp]
\begin{center}
\begin{tabular}{c|c}
\begin{tabular}{ccccccc}
UPO  &  $T$  &   $s$  &  $m$  &  $\Lambda^{-1}$  &  $N$  & frequency \\
\hline \\
 1 & 2.354 & 0.042 & 0 & 6.92E-3 & 12 & high\\
 2 & 5.096 & 0.170 & 0 & 3.83E-3& 10 \\ 
 3 & 13.862 & 6.090 & 0 & 2.41E-3& 6& 2 \\ 
 4 & 14.003 & 5.544 & 3 & 8.66E-5& 6& \\ 
 5 & 14.216 & 5.383 & 3 & 1.07E-4 & 8& \\ 
 6 & 14.316 & 0.132 & 0 & 8.58E-5& 8&\\ 
 7 & 14.578 & 5.918 & 0 & 1.73E-5 & 6& \\ 
 8 & 14.724 & 5.992& 0 & 5.27E-5 & 8&\\ 
 9 & 14.776 & 0.295 & 0 & 4.66E-5 & 7& \\ 
 10 & 14.807 & 0.132 & 0 & 2.08E-5 & 8&\\ 
 11 & 15.030 & 0.758 & 1 & 2.69E-4 & 6 & \\ 
 12 & 15.197 & 5.954 & 0 & 1.58E-6 & 8&\\
 13 & 15.537 & 5.829 & 3 & 1.42E-4 & 8& \\ 
 14 & 16.040 & 0.422 & 0 & 5.28E-5 & 7&2 \\ 
 15 & 16.158 & 0.831 & 1 & 3.05E-6 & 9&\\ 
 16 & 16.667 & 0 & 0 & 7.45E-3 &7& 4 \\ 
 17 & 16.753 & 0.482 & 0 & 4.77E-5 & 10&\\ 
 18 & 16.806 & 0.032& 0 &1.11E-2 & 6&4 \\ 
 19 & 16.899 & -0.018& 0 & 6.39E-3&  7&\\ 
 20 & 16.908 & 0.553 & 0 & 1.31E-5 &9& \\ 
 21 & 16.987 & 0.657 & 3 & 6.85E-6 & 6&\\ 
 22 & 17.160 & 0.361 & 0 & 6.26E-5 & 8 & 6 \\
 23 & 17.204 & 0.526 & 2 & 5.51E-5 & 8&\\
 24 & 17.209 & 0.214 & 0 & 1.74E-4 & 6 & \\
 25 & 17.316 & 0.468 & 0 & 1.15E-5 & 10 & \\
 26 & 17.380 & 0.793 & 2 & 4.04E-7 &  9 & \\
 27 & 17.541 & 5.765 & 0 & 1.16E-6& 8 & \\
 28 & 17.574 & 5.499 & 3 & 2.10E-5 & 9 &\\ 
 29 & 17.848 & 0.646 & 0 & 6.88E-7 & 7&\\ 
 30 & 18.159 & 0.589 & 3 & 4.85E-6 & 7&\\ 
 31 & 18.462 & 0.108 & 3 & 2.75E-4 & 9&\\ 
 32 & 18.694 & 0.434 & 0 & 9.79E-6 & 7&\\
 33 & 18.708 & 0.434 & 0 & 1.41E-5 & 7&\\ 
 34 & 18.878 & 0.418 & 0 & 1.72E-6 & 7 &\\ 
 35 & 18.912 & 5.576 & 0 & 3.36E-8 & 10 &\\ 
 36 & 19.223 & 0.913 & 3 & 7.63E-6 & 6 &\\ 
 37 & 19.334 & 0.375 & 0 & 6.26E-9 & 12&\\
 38 & 19.392 & 5.864 & 0 & 2.06E-10 & 9&\\ 
 39 & 19.402 & 5.957 & 0 & 4.85E-8 & 10 &\\ 
 40 & 19.440 & 5.994 & 3 & 1.37E-5 & 9&2 \\ 
 41 & 19.670 & 5.976 & 0 & 3.78E-4 & 6 &\\ 
 42 & 20.359 & 0.622 & 2 & 7.13E-5 & 7&\\ 
  43 & 20.396 & 0.083 & 3 & 1.75E-6 & 8&\\ 
 & & & & & & \\
\end{tabular}
&
\begin{tabular}{ccccccc}
UPO  &  $T$  &   $s$  &  $m$  &  $\Lambda^{-1}$  &  $N$ & frequency \\
 \hline \\
 44 & 20.612 & 0 & 0 & 1.24E-2 & 8& \\ 
 45 & 20.666 & 0 & 0 & 7.43E-2 & 3&4\\ 
46 & 20.727 & 5.582 & 0 & 5.92E-9& 8&\\ 
 47 & 20.761 & 0 & 0 & 2.24E-2 & 5 &2\\ 
 48 & 20.941 & 0 & 0 & 4.41E-3 & 6 &4 \\ 
 49 & 21.227 & 0.794 & 3 & 5.99E-7 &6 & \\ 
 50 & 21.386 & 6.073 & 0 & 1.02E-3 & 8&2 \\ 
 51 & 21.493 & 5.669 & 0 & 5.69E-7 &7&\\ 
 52 & 21.575 & -0.080 & 0 & 1.29E-4& 6 &2 \\ 
 53 & 21.593 & 0.286 &2 & 3.96E-9 & 11 & \\ 
 54 & 21.854 & 5.954 & 0 & 8.85E-5& 7&\\ 
 55 & 23.243 & 6.120 & 1 & 5.36E-4 & 9&\\ 
 56 & 23.517 & 0.006 & 3 & 5.91E-5 & 7 &\\ 
 57 & 23.551 & 0.034 & 3 & 7.38E-5& 9 &\\ 
 58 & 23.732 & 0.893 & 2 & 2.3E-6 & 6& \\ 
 59 & 24.218 & 0.915 & 2 & 1.56E-4 & 5&\\ 
 60 & 25.136 & 0.028 & 0 & 1.73E-3 & 4&\\ 
 61 & 25.137 & 0 & 0 & 1.53E-3 & 7 & \\ 
 62 & 25.313 & 0.008 & 0 & 6.17E-3 & 8&2 \\ 
 63 & 25.928 & 0 & 0& 3.77E-4& 4&\\
 64 & 26.065 & 0.039 & 0 & 1.49E-4 & 7&\\ 
 65 & 26.329 & 0.159 & 0 & 7.56E-4 & 6&\\ 
 66 & 26.941 & 0.411 & 0 & 1.59E-7 & 9& \\ 
 67 & 28.776 & 6.064 & 2 & 1.94E-7 &6 & \\ 
 68 & 30.147 & 0.212 & 0 & 3.87E-5 & 8&\\ 
 69 & 30.370 & -0.034 & 0 & 3.93E-5& 8 & \\ 
 70 & 30.390 & 0.204 & 0 & 2.28E-4 & 7&\\ 
 71 & 30.473 & -0.185 & 0 & 4.92E-5&6&\\ 
 72 & 30.801 & 0.479 & 0 & 1.19E-5 & 8& \\ 
 73 & 30.808 & -0.43 & 0 & 2.48E-6 & 7&\\ 
 74 & 31.196 & 1.049 & 1 & 1.07E-6 & 5&\\ 
 75 & 32.005 & 5.940 & 0 & 6.13E-6 & 8 &\\ 
 76 & 33.098 & 0.206 & 1 & 2.90E-7 & 6&\\ 
 77 & 34.285 & 0.016 & 0 & 1.14E-5 & 5 &3 \\ 
 78 & 37.494 & 5.807 & 0 & 1.03E-7 & 7&\\ 
 79 & 38.450 & 5.977 & 0 & 1.67E-5 & 8&\\ 
 80 & 40.078 & 5.984 & 0 & 1.01E-5 &6& \\ 
 81 & 50.330 & 6.049 & 0 & 5.97E-6 & 7&2\\ 
 \hline \\
R7 & 2.472 & 0.036 & 0 & 1.05E-1 &9 & high \\
R8 & 1.638 & 0.022 & 0 & 8.61E-3 &14 & high\\
R56 & 16.326 & 0.588 & 2 & 4.81E-5 &6 &\\
R57 & 17.909 & 5.802 & 0 & 5.48E-7&7 & \\
R58 & 20.546 & 0.659 & 2 & 1.90E-5 &8
\end{tabular}
\end{tabular}
\caption{\label{UPOs} Table cataloguing the unstable recurrent flows extracted at $Re=60$ Kolmogorov flow $n=4$, $\alpha=1$. Table shows period $T$, relative $x$-translation shift $s$, $\mathcal{S}$ discrete y-shift $m$, inverse stability coefficient ($\Lambda^{-1}$, equation (\ref{stabcoeff})), the number of unstable directions $N$, and the frequency with which the solution was converged (high is more than 20 and no frequency corresponds to a single convergence). Those prefixed with an `R' are those already found in \citep{Chandler:2013}}.
\end{center}
\end{table}

\bibliography{papers}

\end{document}